\documentclass[
amsmath,amssymb,
aps, prl,
twocolumn
]{revtex4-2}

\usepackage{graphicx}
\usepackage{dcolumn}
\usepackage{bm}
\usepackage[dvipsnames]{xcolor}

\begin{document}
	
	
	
	\def\beq{\begin{eqnarray}}
		\def\eeq{\end{eqnarray}}
	\def\be{\begin{equation}}
		\def\bel{\begin{equation}\label}
			\def\beel{\begin{eqnarray}\label}
				\def\ee{\end{equation}}
			\def\eq{&=&}
			\def\ct{\cite}
			\def\l{\left}
			\def\r{\right}
			\def\bm{\begin{math}}
				\def\me{\end{math}}
			\def\bi{\bibitem}
			\def\om{\omega}
			\def\lt{L(t)}
			\def\al{a(L,T)}
			\def\hf{\frac{1}{2}}
			\def\vr{\vec r}
			\def\ap{\alpha}
			\def\ap{\alpha}
			\def\la{\langle}
			\def\ra{\rangle}
			\def\lbr{\left [}
			\def\rbr{\right ]}
			\def\del{\partial}
			\def\grad{\nabla}
			\def\ul{\underline}
			\def\etal{{\it et al.}}
			\def\lra{\leftrightarrow}
			\def\rar{\rightarrow}
			\def\lb{\label}
			\def\q{\quad}
			\def\qq{\qquad}
			\def\d{\delta}
			\def\D{\Delta}
			\newcommand{\Fig}[1]{Fig.~\textup{\ref{#1}}}
			\newcommand \nn {\nonumber}
			\newcommand \bei {\begin{itemize}}
				\newcommand \eei  {\end{itemize}}
			\newcommand \ii    {\item}
			\newcommand \nt   {\nonumber \\ }
			\newcommand{\fett}[1]{{\mbox{\boldmath$#1$}}}
			\newcommand{\vc}[1]{{\mbox{\boldmath$#1$}}}

			\bibliographystyle{apsrev}
			
\title{Atomistic mechanism of friction force independence on the normal load and other friction laws for dynamic structural superlubricity
			}
			\author{Nikolay V. Brilliantov $^{1,2}$}
			\author{Alexey A. Tsukanov$^{1,3}$ }
			\author{Artem K. Grebenko$^{4}$}
			\author{Albert G. Nasibulin$^{1}$}
			\author{Igor Ostanin$^{5}$}
            \affiliation{$^{1}$ Skolkovo Institute of Science and Technology, 121205, Moscow, Russia}
			\affiliation{$^{2}$Department of Mathematics, University of Leicester, Leicester LE1 7RH, United Kingdom}
			\affiliation{$^{3}$Research and Development Center, TerraVox Global, Paphos, Cyprus}
			\affiliation{$^{3}$National University of Singapore, Singapore}
			\affiliation{$^{5}$University of Twente, Enschede, the Netherlands}
			
			\date{\today}
			
\begin{abstract}
We explore dynamic structural superlubricity for the case of a relatively large contact area, where the friction force is proportional
to the area (exceeding $\sim 100\,nm^2$) experimentally, numerically, and
theoretically.
We use a setup comprised of two molecular smooth incommensurate surfaces -- graphene-covered tip and substrate. The experiments and MD simulations demonstrate independence of the friction force on the normal load, for a wide range of normal loads and relative surface velocities. We propose an atomistic mechanism of this phenomenon, associated with synchronic out-of-plane surface fluctuations of thermal origin, and confirm it by numerical experiments. Based on this mechanism, we develop a theory for this type of superlubricity and show that friction force  increases linearly with increasing temperature and relative velocity, for velocities, larger than a threshold velocity. The MD results are in a fair agreement with predictions of the theory.
			\end{abstract}
			
			\maketitle
			
{\it Introduction. }
Understanding physical nature of friction at different scales, including nanoscales, and ability of controlling it, is of immense fundamental and practical importance \cite{Popov,Drumond}. Indeed, nearly a quarter of irreversible energy losses of today's world industry is attributed to friction \cite{Holmberg2017, Sayfidinov2018}. Therefore superlubricity -- the ultralow friction, due to mutual cancellations of tangential forces for incommensurate surfaces \cite{Vanossi2013,Sankar2020}, seems to be very promising for future technical applications.
It has been first predicted theoretically \cite{Shinjo1990, Shinjo1993} and then confirmed experimentally, e.g. \cite{Dienwiebel2004,SupLub_SurSci2005,Science2015,SupLub_PRL2012,SupLub_PRB2016,Urbah_NaMat2018}, and numerically, e.g.  \cite{SupLub_SciRep2017,Urbah_NaMat2018,Urbakh2019}. 	

Superlubricity has been reported for many nanoscale and microscale systems, ranging from junctions of multilayer graphene flakes and graphite surface, graphene or graphitic junctions, to graphene/graphite-boron nitride heterojunctions \cite{Dienwiebel2004,SupLub_SurSci2005,Science2015,SupLub_PRL2012,SupLub_PRB2016,SupLub_SciRep2017,Urbah_NaMat2018,Urbakh2019}. There exists, however, a number of effects which restricts superlubricity. Among these are incomplete cancellation of tangential forces due to incomplete unit cells of the arising moire pattern at the rim area of the layer (finite size effects), as well as incomplete cancellation within complete unit cells \cite{a,c,d}, atomic scale defects \cite{e} and motion of domain walls in superstructures with large commensurate domains \cite{b}. These effects may give rise to static and dynamic friction. Moreover, superlubricity may be destroyed \cite{SupLub_SciRep2017} by spontaneous variation of surface orientation, resulting in a commensurate state \cite{Urb_PRL_2009}, owing to load-induced commensuration \cite{Kim_PRB_2009,SupLub_SciRep2017}, or other similar effects, e.g. \cite{Urb_PRL_2015}.
			
Most of the above restrictions may be, in principle, surmounted by improving technology, e.g. by diminishing atomic scale defects, decreasing role of the rim area, by increasing the contact size, as demonstrated in Ref. \cite{d}. Still there exists a  restriction for dynamic superlubricity, which remains even for an ideal case of complete incommensarubility and negligible role of finite-size effects. It steams from unavoidable corrugation of the contacting surfaces due to out-of-plane thermal fluctuations of the surfaces.  The importance of such surface deformation for friction, has been demonstrated in \cite{Ouyang2016} for a coarse-grained model of atomic graphene film, sandwiched between two metal surfaces.

Here we address the dynamic friction in structurally superlubric systems for contacting incommensurate surfaces, due to corrugation of the surfaces by out-of-plane thermal fluctuations. We consider the case of relatively large incommensurate contacts, when friction due to rim effects, restricting superlubricity, is small as compared to friction due to  out-of-plane fluctuations. This corresponds  to the "soliton-like, smooth sliding" regime, according to classification of Ref. \cite{d}. Hence, it differs from the most of the studies of dynamic superlubricity, mainly focused on "coherent stick-slip" or "collective stick-slip" regimes \cite{d}, which refer to relatively small contacts. Physically, the addressed mechanism is similar to the one, proposed for commensurate molecularly smooth surfaces of two concentric carbon nanotubes, performing relative telescopic motion  \cite{f,g}.

We investigate such systems experimentally, and by large-scale molecular dynamic (MD) simulations.  In difference to previous studies,  we mimic the  experimental setup using the atomistic (and not coarse-grained) model, along with the most realistic, recent inter-atomic potentials. Also, we develop a theory of such kind of dynamic superlubricity.  In contrast to the previous theories, which assume the friction mechanisms, associated with  Prandtl-Tomlinson (PT) model with thermal activation, e.g. \cite{Jinesh2020,Urb_PRL_2009,PRBKrylov2008,Vanossi2013,SpeedStickSlip} or Frenkel-Kontorova-Tomlinson (FKT) model, e.g.  \cite{Shinjo1993,2DFK2016}, we propose the mechanism of synchronic out-of-plane fluctuations (see below) and confirm it numerically.  Based on this mechanism we explain: (i) the observed independence of the friction force on the normal load and other friction laws, such as (ii) the increase of the friction force with temperature and relative velocity of contacting surfaces (linear above some velocity threshold) and (iii) a linear proportionality of the force with the true contact area.
Although friction independence on the normal load  has been reported \cite{SupLub_PRB2016,Urbah_NaMat2018,lee_comparison_2009} (for somewhat smaller range of parameters), as well as proportionality of the friction force to the contact area \cite{Urbah_NaMat2018} and its increase with temperature \cite{Urbakh2019}, the respective  theoretical description of  all these laws, as a consequence of a specific mechanism, has not been given.

{\it Experimental Results.}
We performed experiments by means of  lateral force microscopy (LFM) -- the regime of the  atomic force microscopy (AFM) designed to explore frictional phenomena at the nanoscale. This method has been successfully utilized before, including the case of graphene over graphene friction studies that support our findings  \cite{berman_nanoscale_2015,lin_friction_2011,lee_comparison_2009}. The measurements were performed in $N_2$ atmosphere with the content of $O_2$ and $H_2O$ less than 1ppm. A typical custom multilayer (Si-Ta-Cu/Si-Pd) probe with graphene on the top is shown in Fig. \ref{fig:lfm}a. For technical details regarding the fabrication and measurements,  see  Supplementary Material (SM), which includes Refs. \cite{hum1,hum2,hum3,hum4,lammps,Feynman,BrilFarad,Timoshenko}.   In short, we utilized a metallic substrate covered with a CVD \cite{grebenko_high-quality_nodate} monolayer graphene and silicon probe covered by metal and graphene synthesized via the same CVD technique; we estimate the contact area of about $10^2-10^3\, nm^2$ (see SM).

The experimental results presented in  Figs. \ref{fig:lfm}b-d, demonstrate very low friction corresponding to structural  superlubricity of incommensurate contact.  They clearly indicate, that within the accuracy of our measurements, the lateral friction force $F_{\rm f}$ does not depend on the normal force, $F_{\rm N}$, up to several $\mu$N, until the graphene coverage of the tip remains stable (see also the discussion in SM).
The friction force increases with the  velocity of the tip, up to $V_{\rm tip} \sim 100\, \mu m \, s^{-1}$, much faster than logarithmically, but somewhat slower than linearly,  see Fig. \ref{fig:lfm}d. Note   that  the velocities in Ref.  \cite{Urbah_NaMat2018}, where the logarithmic dependence of $F_{\rm f}$ on $V_{\rm tip}$ was observed,  were two orders or magnitude smaller. From our results we conjecture that the static friction is vanishingly small.
\begin{figure}[!ht]
				\centering
				\includegraphics[width=\linewidth]{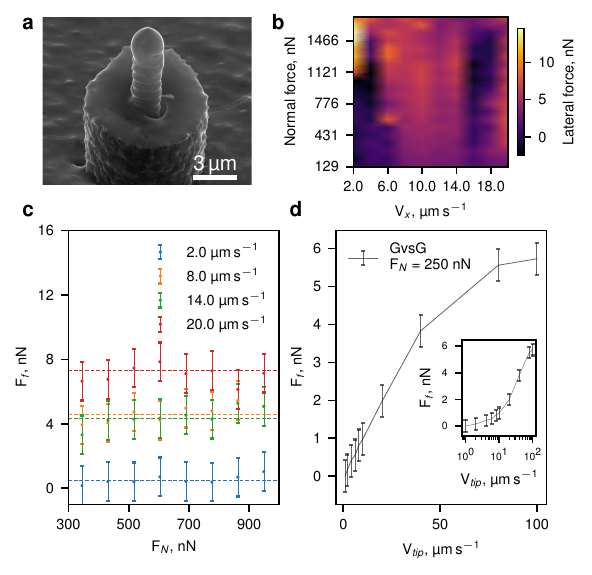}
				\caption{Friction
 experiments by means of AFM: (a) SEM image of the AFM probe covered with copper-graphene composite. (b) Lateral force map illustrating its dependence on the normal force ($F_{\rm N}$) and tip's velocity ($V_{\rm tip}$). (c) Friction force, $F_{\rm f}$, versus normal load,  $F_{\rm N}$, for different tip velocities
 (d) The dependence of the friction force on the tip's velocity;  the inset  shows $F_{\rm f} (V_{\rm tip})$ with logariphmic velocity scale.
				}
				\label{fig:lfm}
\end{figure}
			
{\it MD simulations.}
We performed numerical experiments for the model depicted in  Fig.~\ref{fig:md}a, which mimics the above experimental setup, up to the presence of $N_2$ atmosphere (we assume that its impact is negligible).  Its  bottom part is a planar graphene nanosheet adhered on the surface (111) of  Cu (copper) substrate. The upper part, which models the tip, is a spherical fragment of copper of the initial radius of 300$A^o$; it is  coated by a circular piece of  graphene with the radius of about 100$A^o$. We use incommensurate orientation of the two surfaces; the contact area always exceeded $90\,nm^2$ (see SM). We varied the normal load $F_N$ by three orders of magnitude, from $0.008$~nN to $7.654$~nN. The tip was pulled with a  constant lateral velocity $V_{x}=V$, varying from 0.1 to 5$A^o$/ps.  This range of sliding velocities has been chosen to guarantee the acceptable simulation accuracy. Smaller velocities yielded too noisy data for the friction force, while the thermostating lost stability for larger velocities. The molecular dynamic (MD) simulations have been performed for three different temperatures, T=320~K, 470~K and 670~K.
			
The interactions for graphene were modeled with the use of both, the second-generation REBO potential \cite{REBO2} for intralayer C-C interaction, as well as the (refined) Kolmogorov-Crespi potential \cite{KC2018Urbakh, kolmogorov2005registry} for inter-layer C-C interactions between two different layers. For copper we used the embedded atom method (EAM) \cite{daw1983eam, daw1984eam}, with the  potential developed in \cite{mishin2001structural}. For C-Cu interaction the Abell-Tersoff potential, derived for graphene on Cu substrate has been employed \cite{sule2014CCu}. The horizontal dimensions of the computational cell were $229.88\times 193.90\,A^o$ (in $x$ and $y$ axis), and about $85\,A^o$ in height ($z$-axis). The periodic boundary conditions have been applied along $x$ and $y$ axis.  The total number of atoms was 261064. The bottom half of the substrate and upper half of the tip were coupled to thermostats with temperature $T$; this yields constant $T$ and unperturbed thermal fluctuations at the surface (see SM). The friction force was computed as a time-averaged $x$-component of the total force acting on the upper part of the system.  We have checked that the graphene layers were firmly attached to the according substrates,  and formed together a joint  solid  body. For more computational detail see SM.

\begin{figure}[!ht]
				\centering
				\includegraphics[width=0.9 \linewidth]{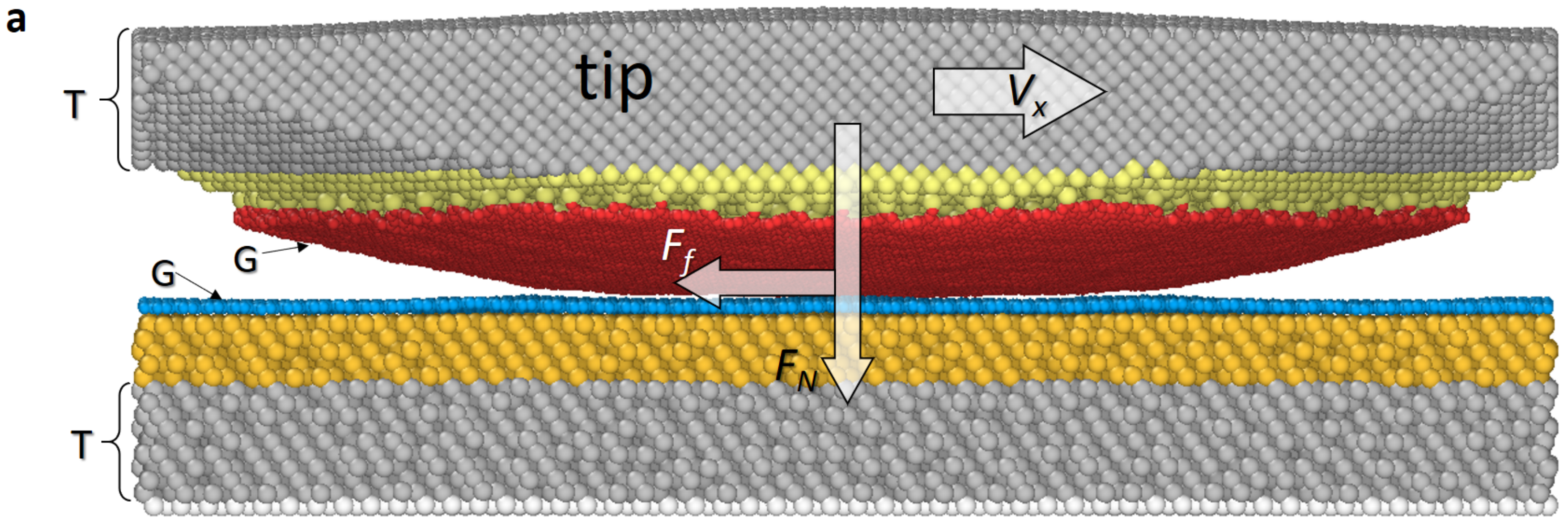} \\
				\includegraphics[width=0.95 \linewidth]{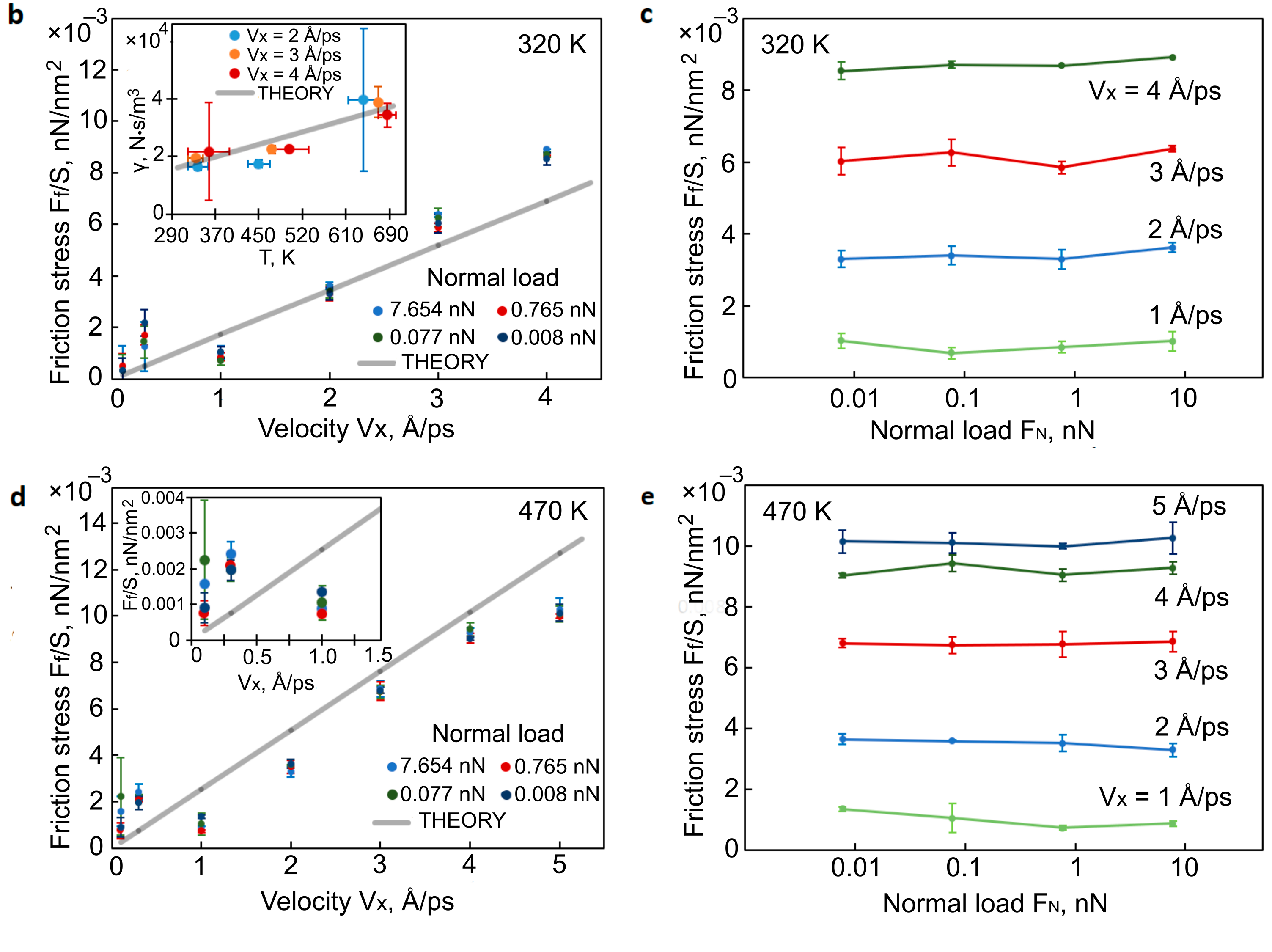}
				\caption{The setup and results of MD simulations. (a)
Graphene nanosheet (shown blue) is firmly adhered on the surface (111) of Cu substrate (orange and gray) and (shown red) on the copper tip (light yellow, gray and white). The groups of atoms coupled to the thermostats of temperature $T$ are given in gray.  The positions of Cu atoms of the bottom substrate  layer (white) are fixed. (b) The dependence of the frictional stress, $F_{\rm f}/S$, where $S$ is the area of the contact, on the tip's velocity $V_x$ for the different normal load $F_N$ at temperature $T \simeq 320$~K. The theoretical dependence, $F_{\rm f}/S= \gamma V_x$, with $\gamma$ given by Eq. \eqref{fin} is shown by thick grey line. The single fitting parameter $\left(b{\eta}/{Y} \right) \simeq (8.5\pm1.5)\cdot 10^{-16}$~s is used, which yields the ratio $(\gamma/T) \simeq 54.0$~${\rm N\, s/(K\,m^3)}$.  The Inset shows the dependence of the friction coefficient, $\gamma=F_{\rm f}/(S V_x)$, on temperature for the normal load of $F_N=0.765$~nN. (c) Frictional stress $F_{\rm f}/S$ as the function of the normal load $F_N$ for different tip velocities at 320~K. (d) and (e) The same as for the main panel (b) and (c) respectively, but for $T \simeq 470$~K.
				}
				\label{fig:md}
\end{figure}
The simulation results are presented in Figs. \ref{fig:md}b -- e.  As it may be seen from the figure,  the MD results confirm the experimental observation of practical independence of frictional force on the normal load. Note that the ``true'', atomic contact area, $S$, remains almost constant for the studied range of loads,  slightly increasing for the largest load of $7.654$~nN; $S$ however noticeably changes with temperature, see SM. Figs. 2c and e depict the frictional stress, $F_{\rm f}/S$ -- the force divided by the contact area. Its practical constancy is consistent with the assumption that friction force is proportional to the contact area, which is large enough (see SM). Moreover, the behavior of the friction coefficient, $\gamma=F_{\rm f}/(S V_x)$ in Fig. \ref{fig:md}b (inset) additionally supports this assumption, see also SM. Our simulations also  confirm the lack of static friction. We do not compare the MD and experimental dependence of the friction force on the velocity, since  the range of experimental and simulated velocities significantly differ (note that $A^o$/ps=100~m/s).
Simulations  demonstrate a linear dependence of friction force on the velocity, above some threshold velocity, see Figs. \ref{fig:md}b and d. The increase of the friction coefficient $\gamma$  with increasing temperature $T$  (see Inset in Fig. \ref{fig:md}b) indicates thermal mechanism of friction, see  below and SM.
			
{\it Theory.}
Both experimental and simulation results, evidence the thermal origin of the friction force -- while surface corrugation, caused by the atomic potential, is very small for incommensurate contact, thermal fluctuations may cause much larger corrugation, hindering the relative motion of bodies at a contact. Here we propose a mechanism which explains friction independence on the normal load. It also explains the velocity and temperature dependence of friction. In contrast to previous studies, which analysed the role of thermal fluctuations in the context of Prandl-Tomlinson model \cite{Vanossi2013,VelDepPRL2000,VelDepTrib2021,chenVelDep2006}, we demonstrate that the major role for molecular smooth, incommensurate surfaces play surface fluctuations of a special type, when surfaces remain in a tight contact; we call them ``synchronic fluctuations'', see Fig. \ref{fig:sketch}. The energy of such fluctuations is significantly smaller than that of other surface fluctuations.  Hence the  synchronic fluctuations, generated by thermal noise, can develop relatively large amplitudes, that is, they dominate. The respective surface corrugation effectively hinders  the surface sliding, giving rise to the friction force.
		
\begin{figure}[!ht]
				\centering
				\includegraphics[width=\linewidth]{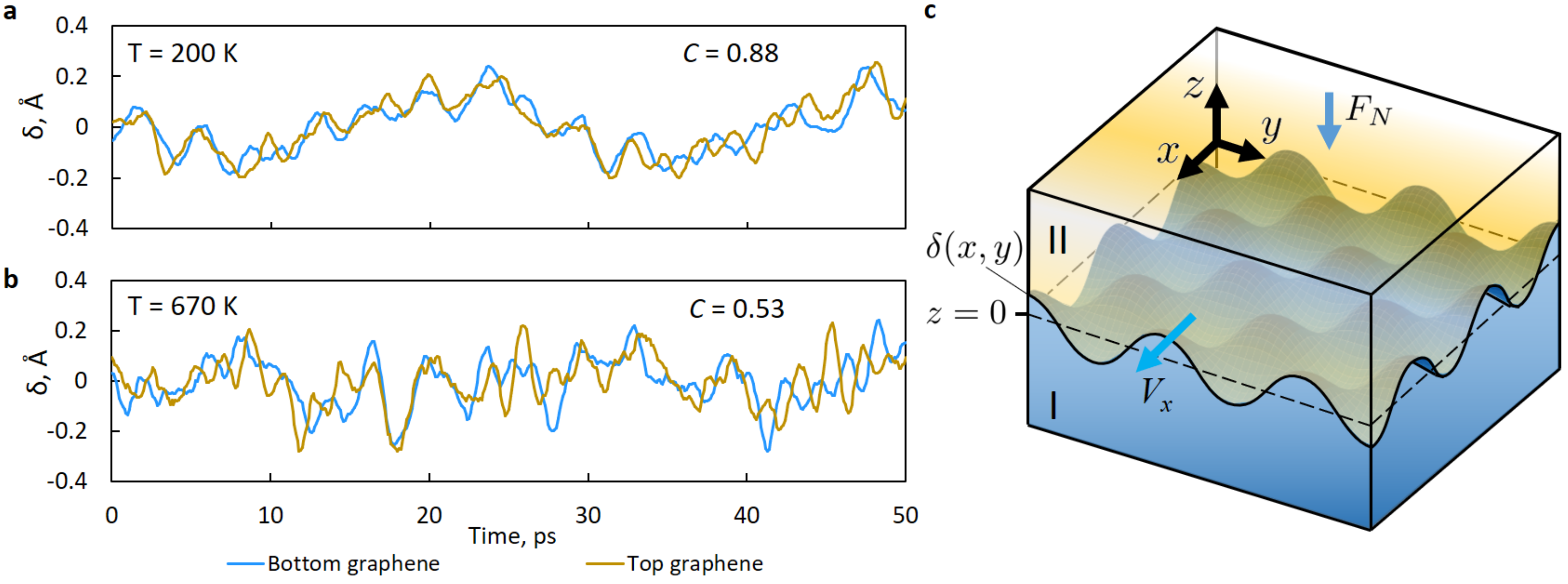}
				\caption{
The synchronicity of surface thermal fluctuations, responsible for friction, demonstrated by MD simulations.  The time dependence of the vertical deviation $\delta(t)$ from the equilibrium plane $z=0$ for the central segment of the bottom (blue) and upper (yellow) surfaces is shown for  200~K (a) and 670~K  (b). The synchronicity is quantified by the correlation coefficient $C$ ($C=1$ for complete synchronicity); it decreases with increasing temperature, see SM for detail. (c) The schematic sketch of the synchronic thermal fluctuations.
}
				\label{fig:sketch}
\end{figure}
To simplify the explanation of the basic mechanism, we consider here an idealized model of a contact in vacuum of two identical flat bodies with a uniform deformation. More rigorous analysis of the general case is presented in SM; it leads to the same conclusion. Let molecular smooth surfaces of such two bodies, with the same elastic and other properties, be in a contact at the plane $z=0$.  Let the bodies be pressed together by the normal load $F_N$, directed along $z$-axis, see Fig. \ref{fig:sketch}. If the area of the contact is $S$, than the normal stress $\sigma_{zz}= F_N/S$ emerges at $z=0$. It  yields the equilibrium deformation of $u_{zz}^{(1)}$ of the first body and $u_{zz}^{(2)}$ of the second body, such that $u_{zz}^{(1)}=u_{zz}^{(2)}= u_{zz}^{\rm eq} =Y^{-1} \sigma_{zz}$, where $Y$ is the Young modulus of the material \cite{Landau:1965}. (Note that the continuum mechanics concepts remain applicable at  the nanoscale, see e.g. \cite{Saitoh}). Then the deformation energy of the two bodies at equilibrium  reads \cite{Landau:1965},
\begin{equation}
				\label{eq1}
				E_{\rm eq} \simeq 2\,  \tfrac12  \int u_{zz}^{\rm eq} \sigma_{zz} d {\bf r}  =   Y  \left(u_{zz}^{eq}\right)^2\,S L,
\end{equation}
where the integration is performed over the whole volume of the bodies, equal to $S\, L$, with $L$ being their dimension in the direction along $z$-axis.  Consider now a  small deviation, $\pm \delta(x,y)$ of the both surfaces from the equilibrium position at $z=0$, such that the surfaces remain in a tight contact. These deviations correspond to the synchronic fluctuations, see Fig. \ref{fig:sketch}.  The respective deformations are:
\begin{equation}
			\label{eq2}
				u_{zz}^{(1)}=u_{zz}^{\rm eq} +\tilde{\delta}(x,y), \qquad \qquad u_{zz}^{(2)}=u_{zz}^{\rm eq} -\tilde{\delta}(x,y) ,
\end{equation}
where  $ \tilde{\delta}(x,y)=\delta(x,y)/L$ is the deformation caused by the surface fluctuation $\delta(x,y)$. The energy of the synchronic fluctuation then reads,
\begin{eqnarray}
				\delta E \!\!&=& \!\!\frac{Y}{2} \!\int  \! d{\bf r}\! \left[\left(u_{zz}^{\rm eq} +\tilde{\delta}(x,y)\right)^2 \!+\! \left(u_{zz}^{\rm eq} -\tilde{\delta}(x,y) \right)^2\right]\! - E_{\rm eq} \nonumber \\
				&= &(Y/L)  \int_{S} dxdy \, \delta^2(x,y),
				\label{eq3}
\end{eqnarray}
that is,  $\delta E = {\cal O}(\delta^2)$, which means that the energy of such  synchronic fluctuations are second order with respect to the amplitude $\delta$. The lack of linear-order  terms in the fluctuations energy, $\delta E$, makes them much more energetically favorable, then non-synchronic fluctuations, which energy contains linear-order terms in $\delta$. Hence, the synchronic fluctuations play a major role in thermal corrugation of the surface, providing  the main friction mechanism for incommensurate surfaces. The dominance of synchronic fluctuations is  directly confirmed by numerical simulations, see Fig. \ref{fig:sketch}.
			
Even more important, is that the fluctuation energy $\delta E$ does not depend on the equilibrium deformation $u_{zz}^{\rm eq}$, that is, it is independent on the normal load, $F_N$. Hence, we come to the principle conclusion -- the friction force (for this type of superlubricity) is mainly determined by the surface corrugation, due to synchronic fluctuations, and does not depend on the normal load.
			
Now we estimate the dependence of the friction force $F_{\rm f}$ on the relative velocity $V$ of the surfaces. We will not discuss here the dependence $F_{\rm f}$ for small velocities, but  will address the velocities larger than a threshold velocity $V_*$, that may be associated with the propagation velocity of  the synchronic  fluctuations along the surface. Based on the estimates detailed in SM, we obtain, $V_* \sim 50 - 100 $\,m/s.
			
Consider the dissipation of energy due to relative motion of two surfaces with the velocity $V >V_*$. We assume that for $V >V_*$, the surfaces, corrugated by the synchronic fluctuations, remain in a tight contact, so that
\begin{equation}
				\label{eq4}
				u_{zz}^{(1/2)}(x,y,t)= u_{zz}^{\rm eq} \pm \tilde{\delta}(x-Vt,y).
\end{equation}
This means that the  sliding  motion ``drives the wrinkles'' of the synchronic fluctuations in the direction of the relative motion, see Fig. \ref{fig:sketch}c. Then the deformation $u_{zz}^{(1/2)}(x,y,t)$ varies in time, as
\begin{equation}
				\label{eq4a}
				\tfrac{d}{dt} {u}_{zz}^{(1)}= -\tfrac{V}{L} \, \tfrac{\partial}{\partial x}   \delta(x,y)  , \qquad  \tfrac{d}{dt} {u}_{zz}^{(2)}= \tfrac{V}{L} \, \tfrac{\partial}{\partial x}   \delta(x,y),
\end{equation}
yielding the dissipation of energy per unit time $W_{\rm diss}$. It is  quantified by the dissipative function $R$  \cite{Landau:1965}.  In our case all deformation components except $u_{zz}$ may be neglected, which yields  (see SM for more detail),
\begin{eqnarray}
				\label{eq5}
				R &=& \eta \left( \frac{d{u}_{zz}^{(1)}}{dt} \right)^2 +
				\eta \left( \frac{d{u}_{zz}^{(2)}}{dt} \right)^2,   \\
				W_{\rm diss} &= & \int R \, d{\bf r}\!  = \left(\frac{2\eta}{L}\right)  \, V^2 \int_S \left( \frac{\partial \delta(x,y) }{\partial x } \right)^2 dx dy .
				\nonumber
\end{eqnarray}
Here $\eta= (\tfrac49 \eta_1 + \tfrac12 \eta_2)$, with $\eta_1$ and $\eta_2$ being the viscosity coefficients of solid material, quantifying viscous losses, respectively,  for shear and bulk deformation rates \cite{Landau:1965}.
			
Since the thermal fluctuations, $\delta(x,y)$, are random, the averaging of $W_{\rm diss}$ is needed. It may be done using the probability  of such fluctuations, $P(\delta) =Z^{-1}e^{-E(\delta)/k_BT} $, where $E\left(\delta (x,y) \right)$ is the energy of the fluctuation $\delta (x,y)$, $T$ is temperature, $k_B$ -- the Boltzmann constant and $Z^{-1}$ -- the normalization factor. The averaging is to be performed over all possible $\delta (x,y)$:
\begin{equation}
				\label{eq6}
				\langle W_{\rm diss} \rangle = \int {\cal D}[\delta (x,y)] \,W_{\rm diss} [\delta (x,y)] \,P[\delta (x,y)].
\end{equation}
${\cal D}[\delta (x,y)] $ in (\ref{eq6}) denotes functional integration over two-dimensional functions, associated with the surface fluctuations at  the contact area $S$. Note that Eq. \eqref{eq6} shows that $\langle W_{\rm diss} \rangle$ does not depend on the normal load. Indeed, $E(\delta)$, and hence of $P(\delta)$ in Eq. \eqref{eq6}, do not depend on $F_{\rm N}$. Referring for computation details to SM (where it is done for the general model), we present the final result:
\begin{equation}
				\label{eq7}
				\langle W_{\rm diss} \rangle = F_{\rm f} \, V = \pi b (k_BT/Y)  \rho^2_s  \eta \,  V^2 S,
\end{equation}
where $\rho_s$ is the number of surface atoms per unit area and we take into account that the dissipation power is equal to the product of the velocity $V$ and friction force $F_{\rm f}$. The numerical coefficient $b$ depends on the Poisson ratio $\nu$ and combination of viscous constants, $r=(\tfrac49 +\tfrac12 \eta_2/\eta_1)^{-1}$ as,
\begin{equation}
				\label{eq8}
				b= \frac{(1+\nu)(1-2\nu)}{\pi^2}\,\left[  \frac{b_1+rb_2}{b_2+(1-\nu)b_1 }\right],
\end{equation}
where we abbreviate,  $b_1\equiv 5-8\nu +8 \nu^2$ and $b_2 \equiv 13-20\nu +8 \nu^2$, see SM. This yields the friction force and coefficient:
\begin{equation}
				\begin{split}
					\label{fin}
					F_{\rm f} &=  \gamma S V, \\   \gamma &= \pi b  (k_BT/Y)\rho_s^2 \eta.
				\end{split}
\end{equation}
Hence, we demonstrate that the  friction force  does not depend on the normal load, in a qualitative agreement with the experiment and simulations, Figs. \ref{fig:lfm} - \ref{fig:md} and explain the atomistic  mechanism of this phenomenon. We also show that it  linearly depends on the sliding velocity $V$ and contact area $S$, while  the friction coefficients  $\gamma$ linearly increases with temperature, in agreement with the numerical experiments,  Figs. \ref{fig:md}b-e.
Noteworthy, the friction coefficient $\gamma$ is expressed in terms of the square average of thermal synchronic fluctuations, $ \gamma \sim \langle W_{\rm diss} \rangle \sim \eta \langle (\partial \delta /\partial x )^2 \rangle$,  that is, it obeys the fluctuation-dissipation relation \cite{resibua,SokolovEbel}; a similar linear dependence on $T$ of the dynamic friction force was reported for double-wall nanotubes \cite{g}.

			{\it Conclusion. }
We explore dynamic structural superlubricity experimentally and numerically, using incommensurate contact of two solid surfaces with firmly adhered graphene layers. The contact area was relatively large, corresponding to ``soliton-like smooth sliding'' regime, by the classification of Ref. \cite{d}, where friction force is proportional to the contact area.  We observe superlubricic behavior for a wide range of the normal load and relative velocities of  surfaces, spanning several orders of magnitude. The impact of temperature has been also investigated.  For this kind of superlubricity we propose a few friction laws   and support them by theoretical analysis. In contrast to conventional Amontons-Coulomb laws of dry friction \cite{Popov,Drumond}, we demonstrate independence of friction force on the normal load  and its increase with increasing inter-surface velocity and temperature. Furthermore, we show that above some threshold velocity the dependence of friction force on the velocity is linear;  the temperature dependence is also linear.  We propose an atomistic mechanism of friction independence on the normal load in terms of synchronic out-of-plane surface fluctuations of thermal origin, when two surfaces remain in a tight contact. Due to relatively small energy, such fluctuations dominate, yielding corrugation of the contacting surfaces, which hinders the relative  motion. We confirm this mechanism in our MD experiments. We show that the intensity of such synchronic fluctuations, which determine friction,  does not depend on the normal load. As the result, friction force does not depend on the normal load, for both -- small velocities (as in the experiment) and large ones (as in MD simulations). For large velocities (exceeding  the propagation velocity of synchronic fluctuations) we develop a theory of the friction force. It is in a fair agreement with  the simulation data. We  show that the friction coefficient obeys the fluctuation-dissipation relation, which is similar to bulk viscous friction but is very unusual for conventional dry friction.

\section{Acknoledgement}
We are thankful for Michael Urbakh for valuable comments and drawing our attention to Refs. \cite{Urb_PRL_2009,Urbakh2019}.

\bibliographystyle{apsrev4-2}
%

\end{document}


\flushbottom
\maketitle

\section{Experiment}
\subsection{Graphene synthesis}
For the friction measurements two types of samples have been fabricated. The first, referred here as a substrate, was the metal catalyst with a large area. It had either the form of a metal foil (Pd) or massive crystal (Cu) covered with the graphene monolayer. These samples exactly followed the protocol described elsewhere\cite{grebenko_high-quality_nodate}. As the result, graphene formed by chemical vapor deposition on the surface of Pd and Cu formed monocrystallines of size exceeding $100\, \mu m$.   
The second was a silicon cantilever (HA\_CNC and HA\_FM, Ostec, Russia, k = 3.5 N/m) covered with a thick metallic layer and graphene. Due to the catalytic decomposition of carbon monoxide over the metallic substrate,  mono- and few-layered graphene has been formed on the surface of the both samples. In order to form graphene on the tip, two methods have been applied. In the first method, the cantilever tips have been covered with 5 nm of chromium and 200 nm of palladium and then subjected to the same synthesis procedure as for the substrate.  For the copper catalyst, the tip was initially subjected to the magnetron deposition of tantalum, acting as a wetting layer for the melting copper on the top of it. Afterwards, the copper film was deposited on the top, using the thermal evaporator Tecuum AG. Both for the deposition and formation of the copper substrate 99.999 \% copper pellets (Kurt Lesker) have been utilized. 

\subsection{Atomic Force Microscopy}
Bruker Multimode V8 atomic force microscope has been  employed for the friction measurements in the regime of lateral force microscopy. For each particular load force,  512x512 scan over 10x10 $\mu m^2$ area has been used. Trace and Retrace signals were captured, subtracted to remove the impact of topography, and additionally filtered to remove the contribution of steps inevitably formed during the graphene synthesis. As a test to check the presence of the graphene on the tip,  we have deliberately removed it. This was done by the deformation of the tip's apex by applying very large load, which resulted in a drastic increase of friction force. The thermal tune, along with PeakForce tapping\textsuperscript{TM} force-distance curve based calibrations, have been used for the precise measurements of the cantilever's spring constant and deflection sensitivity. The tip radius was measured using the dimpled aluminium sample, fabricated at the Moscow State University. The lateral force microscopy data has been processed using pySPM, scypy, numpy and pandas libraries. In the experiments we were limited by about 80 \si{\micro\meter\per\s} for the scanning velocity and by  approximately 1000 \si{\nano\newton} for the load force. 

In our experiments we did not study the  dependence of the friction force  on the tip-substrate orientation, expecting that due to random orientation of the substrate and tip, it is very low probability that it results in a commensurate system.  This is confirmed by the experimental results presented in  Figs. 1b-d, of the main text, demonstrating very low friction. Note that in Fig.1b,  we show the "raw data" -- the friction map, where both velocity and load are varied.  The cross-sectional force traces are shown in Fig. 1c and Fig. 1d. That is, fixing the velocity and varying the normal load (the "vertical" cross-section of the Fig. 1b) yields Fig. 1c, while fixing the normal load and varying the velocity (the "horizontal" cross-section of the Fig. 1b) yields Fig. 1d.

To avoid the effects of capillary condensation, that may influence the dependence of the friction force on the normal force and  velocity, as it has been shown in Refs. \cite{hum1,hum2,hum3,hum4}, we performed AFM measurements in a controlled atmosphere. Namely, the atmospheric hood was installed, sealing AFM head, sample compartment and tip holder. Samples were installed after annealing at $100 \, C^o$, and immediately after that, the  atmospheric hood was purged with $N_2$ with the content of Oxygen and Water less than 1ppm ($[0_2]<1ppm$, $[H_2O] < 1ppm$) at 3 lpm (liter per minute) for 24 hours. During the measurements the $N_2$ flow was reduced to 200 sccm  (standard cubic centimeters per minute) to minimize the influence of the flux on the AFM tip.

\subsection{Samples' characterization}
DXRxi Raman Imaging Microscope with 532 nm excitation laser has been used to determine graphene presence on the surface of metal catalyst. Experiments to determine graphene presence on the AFM tp were performed with the confocal Raman microspectrometer InVia Qontor (Renishaw, UK) with 532 nm laser. Scanning Electron Microscope Jeol JSM-7001F at acceleration voltages from 10 to 30 kV was employed  to analyze the cantilever's tips.

\subsection{Raman Spectroscopy and topography images}

\begin{figure}[ht]
\centering
\includegraphics[width=\linewidth]{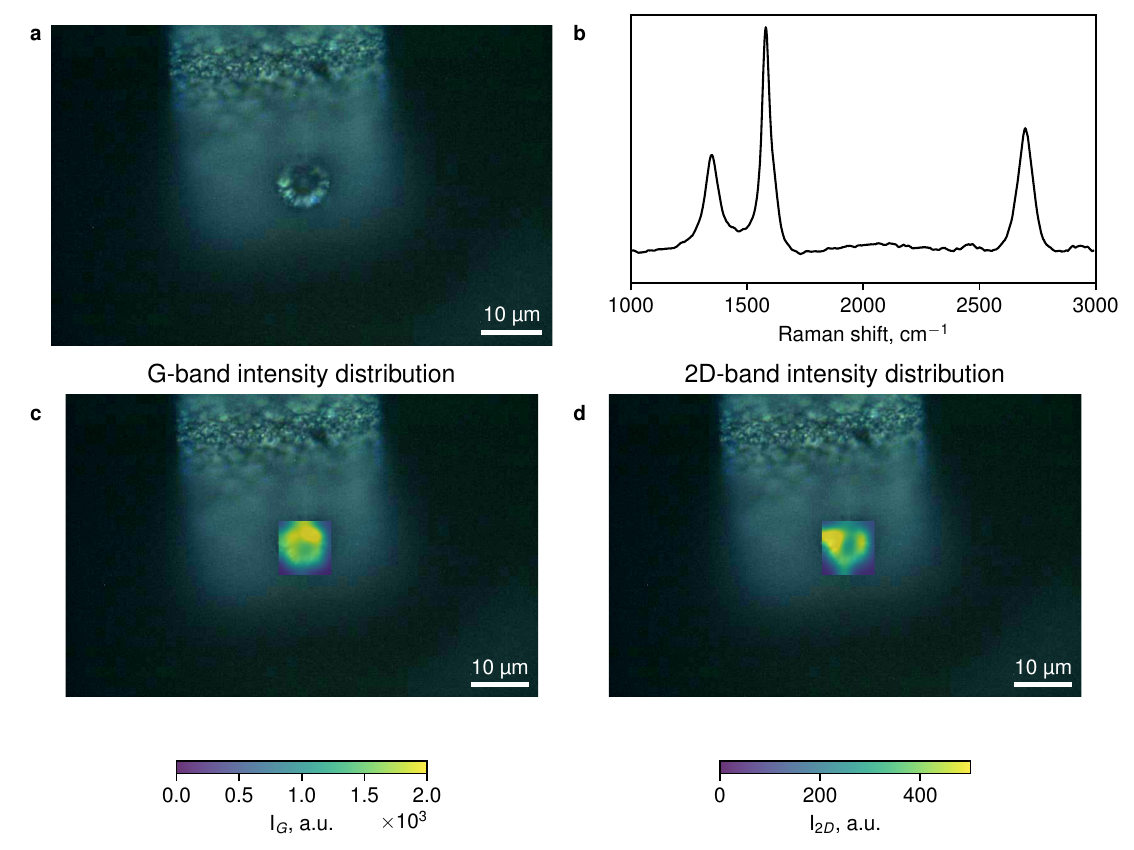}
\caption{\textbf{Raman spectroscopy of graphene covered tip} \textbf{(a)} Optical image of the tip shown in \textbf{Figure 1} of the main text \textbf{(b)} Typical individual Raman spectrum from an arbitrary point from the cantilever chip. \textbf{(c)} Distribution of the hallmark G-band over the tip apex. Raman map is 10x10 points with 1 \si{\micro\metre} step. \textbf{(d)} 2D-band distribution map.}
\label{fig:sup_ram}
\end{figure}

In order to prove the presence of graphitic carbon on the surface of a cantilever we have performed Raman spectroscopy mapping over all probes that we used in lateral force microscopy experiments. Typically, these samples had graphene covering of the  all surface of the chip, including the tip. Due to the developed surface and non-optimal synthesis conditions (lowered synthesis temperature of 900 \si{\celsius} instead of 1085 \si{\celsius} and long cooling times of 90 minutes instead of 10)  the obtained graphene might have different number of layers in different regions. Optical resolution limitations did not allow us to precisely determine the number of layers  at the tip apex. The results are sketched in Figure \ref{fig:sup_ram}; more details will be published in the forthcoming study.   

\begin{figure}[!ht]
				\centering
				\includegraphics[width=0.49 \linewidth]{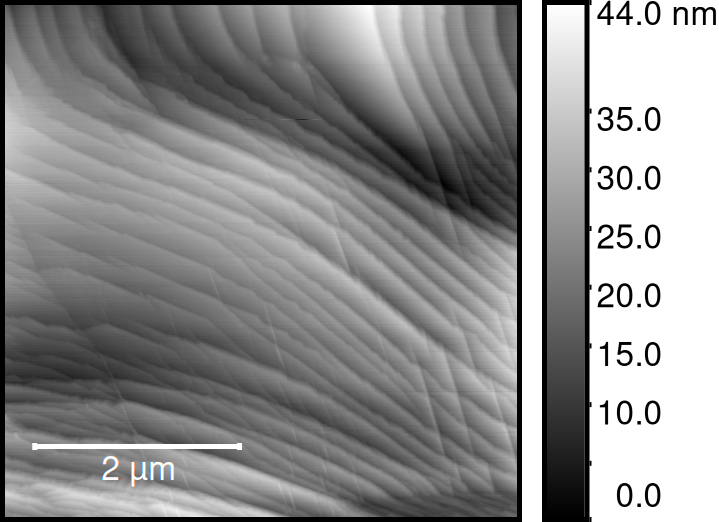} 
                 \includegraphics[width=0.49 \linewidth]{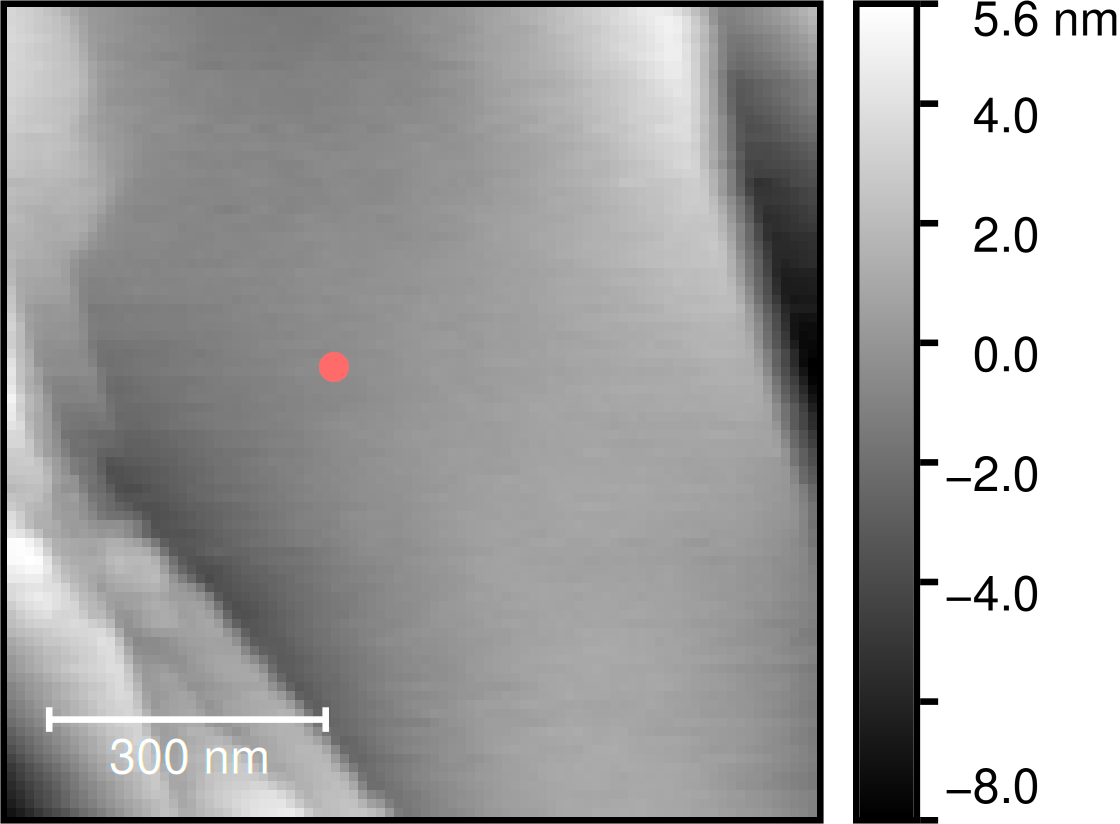} 
				\caption{Left panel: The full topography image of the substrate -- crystalline surface of Cu covered with graphene, used for the measurements of the friction force. Right panel: A small piece of the surface shown in the right panel. The red spot illustrates an approximate size  of the contact of the AFM tip with the substrate.  }
				\label{fig:Top1_2}
\end{figure}

To assess the  surface properties of the substrate,  we exploited  the topography  images of the crystalline surface of the Cu+Graphene, used for measurements of the friction force, see Fig. \ref{fig:Top1_2}.
Here we also show the estimated size of the contact of the AFM tip. As it may be clearly seen from the figure, the surface possesses flat and defect-free domains of size of about a few hundred nanometers. Based on this, we believe that the observed superlubricity arises due to incommensurate, defect free contact, between the tip and substrate. 

To assess an average crystalline size of graphene, $L_D$ (on the tip), we utilize  the Raman spectra collected from the tip. Here we use the respective  relation (see e.g. M.M. Lucchese et al., Carbon, 48 (2010) 1592):
$$
L_D^2(nm^2) =2.4 \times 10^{-9} \, \lambda_L^4 (I_G/I_D)
$$
where $\lambda_L$ is the wavelength of the laser beam of 532 nm, $I_G $ is the G-band intensity and $I_D$ is D-band intensity. In our experiments the ratio $(I_G/I_D)$ was larger than 2.5, which yields the estimate of about $25\, nm$ for the size of defect-free region, corresponding to the area of $\sim 10^3\, nm^2$. This coincides with the expected size of the contact spot in our experiments, with  the area of about $10^2-10^3\, nm^2$ (see below).
Moreover, according to Grebenko et. al, Ref. [32] protrusions and other bulging parts of a surface tend to be centers for nucleation with higher probability as compared to other, flat parts of the surface. Hence, in our case,  when we grow graphene on the cantilever tip, there is a high probability, that the apex of the tip acts as a nucleation center. This results in a defect free area of size of about $25\, nm$ centered at the tip's apex. The existence of the defect and contaminants free area at the tip's apex was further controlled by the friction measurements. 

The contact area may be roughly estimated from the Hertz contact law \cite{Landau:1965}, 
$
F_{\rm N} \sim \sqrt{R} \, Y \xi^{3/2}
$,
where $R$ is the tip radius ($R\sim 10^{-6}\, m$), $Y$ is the Young modulus of the tip material and $\xi$ is the tip deformation. Using $Y\sim 10^{11}-10^{12}\, N/m$ and $F_N \sim 10^{-7}-10^{-6} \,N $, we find $\xi \sim 10^{-10}-10^{-9} \, m$. With the relation for the contact radius  \cite{Landau:1965}, 
$
a^2 \sim R\xi
$, 
we arrive at the estimate of $10^2-10^3 \, nm^2$ for the area of contact. 

\subsection{Sample selection}
The results presented in Fig. 1c of the main text demonstrate both, negative and positive slopes $\mu$  in the dependence, $F_{\rm f} =F_{\rm f,0} + \mu F_{\rm N}$,  of friction force on the normal load.  That is, $\mu$ ranges from $-1\cdot 10^{-4}$ (for $V_{\rm tip}= 2 \, \mu m/s$) to $+8 \cdot 10^{-4}$ (for $V_{\rm tip}= 20\, \mu m/s$), with the maximal $\mu=2 \cdot 10^{-3}$ for (for $V_{\rm tip}= 8\, \mu m/s$). Therefore, following Ref. [19] of the main text, we conclude, that within the accuracy of our measurements friction force does not depend on the normal load. 

Some of the fabricated samples contained graphene monocrystallyne boundaries, as well as contaminants at the region of the tip apex. Such samples demonstrated a large friction force and large slopes $\mu \sim 10^{-2} $  in the dependence of the friction force on the normal load (which it about 100 times larger, than for the case of "appropriate" samples). This is illustrated in Fig. \ref{fig:Defect} below. 
\begin{figure}[!ht]
\centering
\includegraphics[width=0.7\linewidth]{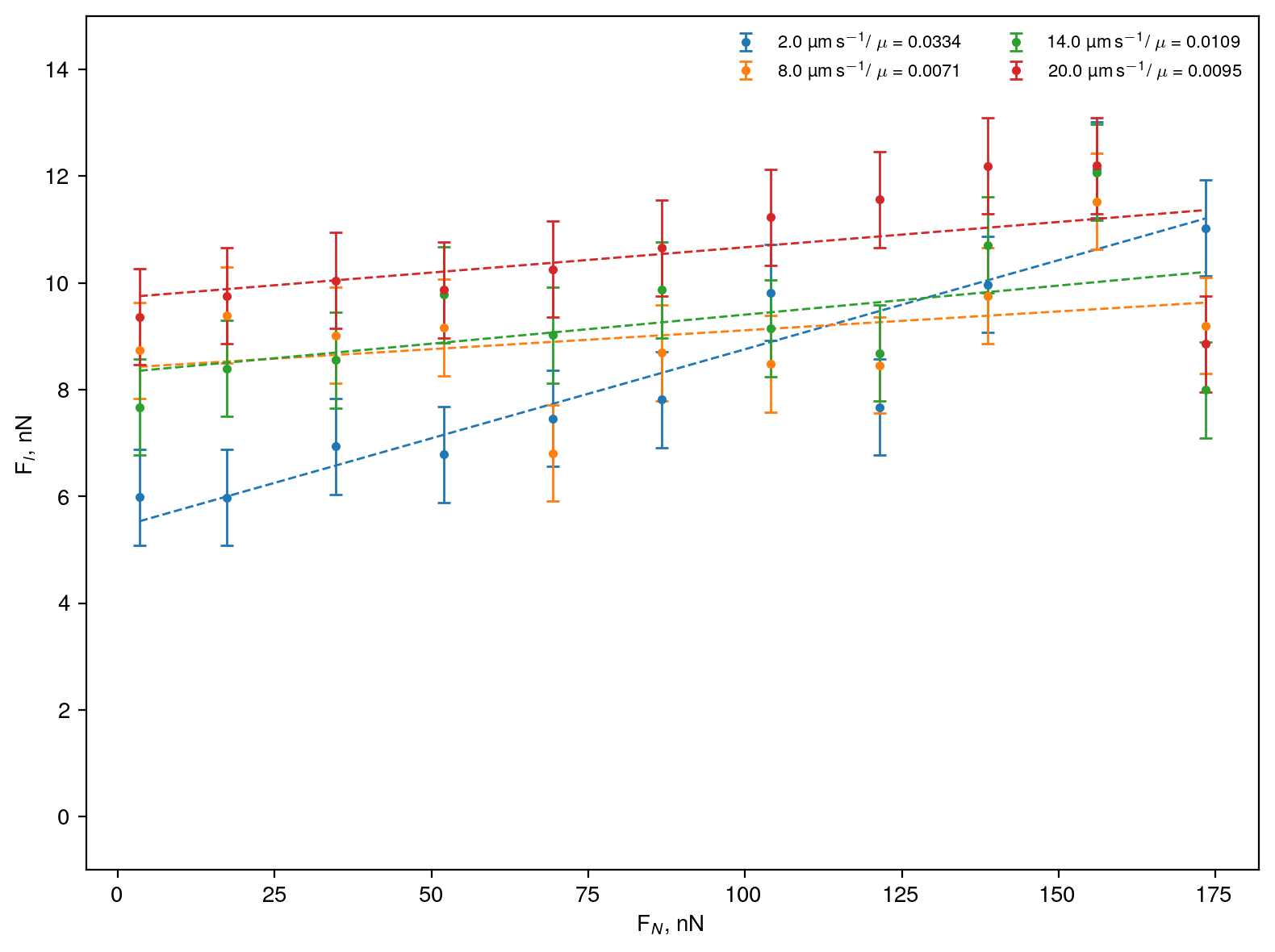}
\caption{Friction force as a function of the normal load for a defective tip, that is, which contains  graphene monocrystallyne boundaries and/or contaminants at the region of the tip apex. This kind of defective tips demonstrate a large friction force and large slope $\mu$ for the dependence, $F_{\rm f} =F_{\rm f,0} + \mu F_{\rm N}$,  of the friction force on the normal load. 
				}
				\label{fig:Defect}
\end{figure}
Such tips were filtered out, as defective, during the preparation of the experimental setup. 

Hence, we have chosen only those tips, that demonstrated an evident superlubrical behavior and filtered out the rest as defective (that is, containing contaminants and/or grain boundaries at  the tip apex).

\section{MD simulations}
We performed the MD simulations using LAMMPS software~\cite{lammps}.  The interaction for graphene was modeled with the use of both, the second-generation REBO potential \cite{REBO2} for intralayer C-C interaction, as well as the (refined) Kolmogorov-Crespi potential \cite{KC2018Urbakh, kolmogorov2005registry} for inter-layer C-C interactions between two different layers. Copper was modeled utilizing the embedded atom method (EAM) \cite{daw1983eam, daw1984eam}, with the  potential developed in \cite{mishin2001structural}. For C-Cu interaction the Abell-Tersoff potential, derived for graphene on Cu substrate has been employed \cite{sule2014CCu}. The horizontal dimensions of the computational domain were $229.88\times 193.90$~{\AA} (in $x$ and $y$ axis), and about 85~{\AA} in height ($z$-axis). The periodic boundary conditions have been applied along $x$ and $y$ axis.  In order to maintain constant temperature for the bottom half of the substrate and the upper half of the tip, two thermostats were deployed with the same temperature $T$. 

Our numerical model corresponds to the system under the UHV conditions, while the real experiments were performed in the nitrogen atmosphere with the content of $O_2$ and $H_2O$ less than 1 ppm. We believe that the presence of such  $N_2$ atmosphere has a negligible impact on the friction force, which will be the same as in the UHV conditions. 

We used the implemented in LAMMPS~\cite{lammps-rescale} thermostatting through the velocities rescaling. This type of thermostat (1) is simple and therefore computationally cheap and (2) doesn't apply any external random forces to atoms, in contrast with, for example, Langevin thermostat. Using Langevin thermostat  at the preliminary stage of our research,  resulted in a significant thermal noise, that made the assessment of the friction force by MD unreliable. 

Owing to the high thermal conductivity of copper, the application of the thermostat only to the half of the tip and half of the substrate guarantees that the real interface possesses the same temperature as the thermostat. At the same time, direct application of the thermostat to atoms at the interface may result in artificial, thermostat-dependent microscopic behavior. Since we explore the atomistic mechanism of superlubricity, we wish to exclude this. Hence applying the thermostat only to the atoms which are  a few layers apart from the contact surface, allows to avoid the thermostat-dependent artificial behavior of  the atoms in the contacting layers and, at the same time, keep there the required temperature. In this way we were able to detect the synchronic out-of-plane thermal fluctuations.

The total number of atoms was 261064, including 118560 Cu atoms of the substrate, 113176 Cu atoms of the tip, 17064 carbon atoms of the bottom graphene nanosheet and  12264 atoms of the upper graphene nanosheet. The complete MD model \emph{will be} available in GitHub via the link~\cite{github-tsukanov}.

\subsection{Friction stress force as a function of the normal load }
In order to compute the friction force in the MD simulations, the x-component of the total force acting on the upper part of the system was computed each 10 time steps. (The total force implies the sum of all forces acting on each atom of the upper part. Note, that these  atoms act on each other, but  all the internal forces cancel, owing to the $3^{\rm rd}$ Newton's law,  resulting in an average friction force for a given velocity). Then we performed the  time averaging. This time-averaged force was our numerical estimate for the friction force.

\begin{figure}[ht]
\centering
\includegraphics[width=0.95\linewidth]{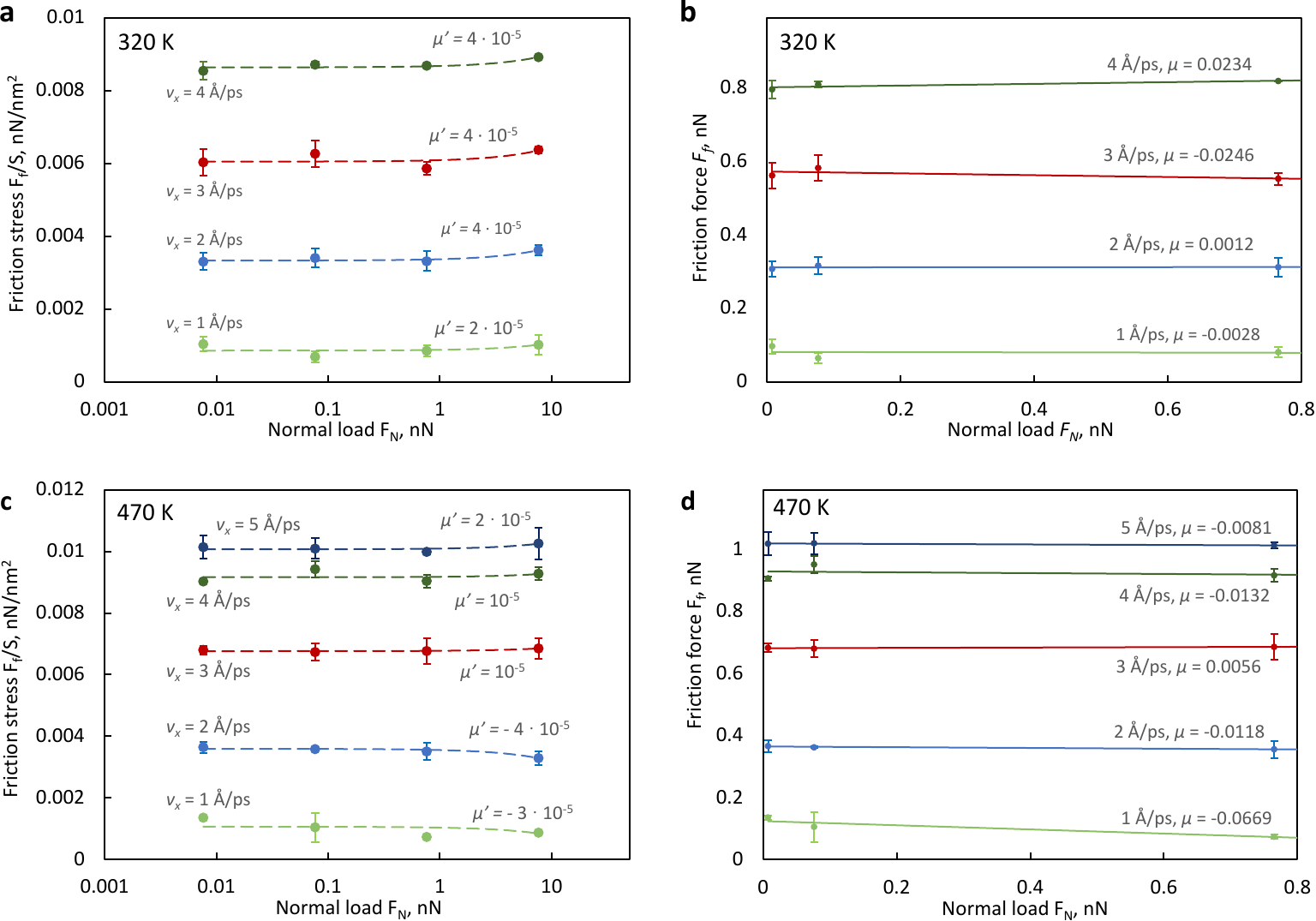}
\caption{\textbf{Dependence of  friction stress and friction force on the normal load from  MD simulations. } The dependence of the friction stress $F_{\rm f}/S$ on the normal load,  ({\bf a, c}). It is quantified by the coefficient $\mu^{\prime}$ in the relation, $(F_{\rm f}/S) = \sigma_0+ \mu^{\prime} F_{\rm N}$. Similarly, the dependence of the friction force itself, on the normal load  ({\bf b, d}), is quantified by $\mu$ in the relation $F_{\rm f}= F_{\rm f,0} + \mu F_{\rm N}$. Note, that both $\mu^{\prime}$ and $\mu$ are very small (and alter their sign), manifesting the independence of the friction stress on the normal load with the accuracy of MD experiments. 
}
\label{fig:md-mu}
\end{figure}

The results of MD simulations showed that the friction stress $F_{\rm f}/S$ is almost independent on the normal load $F_{\rm N}$; this is illustrated in Figs. \ref{fig:md-mu} a,c. Here the slope $\mu^{\prime}$ in the dependence, $(F_{\rm f}/S) = \sigma_0+ \mu^{\prime} F_{\rm N}$, is very small: $\mu^{\prime}$ is of the order of $10^{-5}$ (note that $\mu^{\prime}$ has the dimension of $[1/m^{2}]$ and quantifies the slope of the friction stress in units $[nN/nm^2]$ in terms of normal load in units $[nN]$). The dependence of the friction force itself, on the normal load, $F_{\rm f}= F_{\rm f,0} + \mu F_{\rm N}$ is depicted in Figs. \ref{fig:md-mu} b,d. Here the slope $\mu $ is also small. 

The smallness of $\mu^{\prime}$ and alteration of  its  sign ($\mu$ is also small and alters sign),  leads us to the  conclusion that the friction stress, $F_{\rm f}/S$, does not depend on the normal load, within the accuracy of the MD experiments.   

\subsection{Simulation results for 670~K and temperature dependence of the contact area}

\begin{figure}[ht]
\centering
\includegraphics[width=1.0\linewidth]{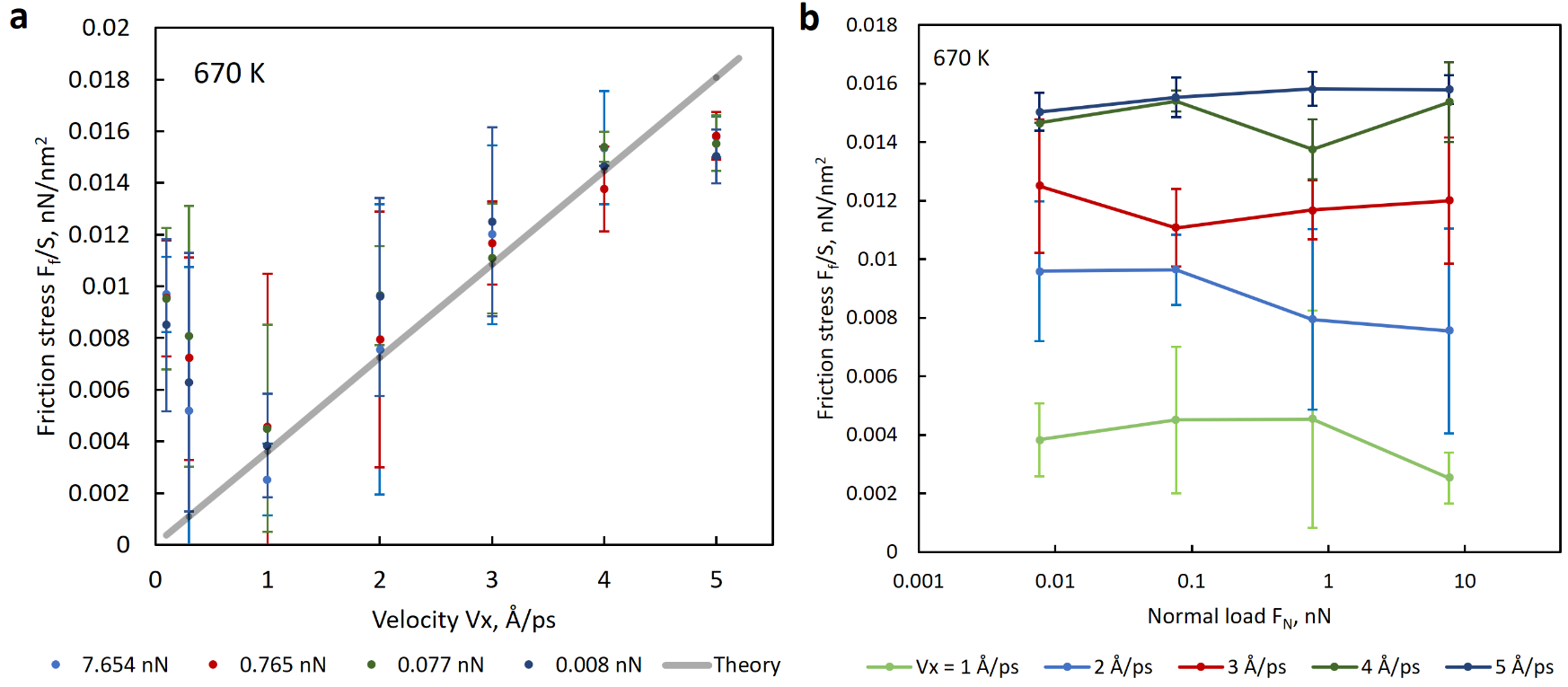}
\caption{\textbf{Results of MD simulations for temperature 670~K.} \textbf{(a)} The dependence of the measured frictional stress, $F_f/S$, where $S$ is the area of the contact, on the tip's velocity $V_x$ for the different normal load $F_N$ at temperature $T \simeq 670$~K. \textbf{(b)} Frictional stress $F_f/S$ as the function of the normal load $F_N$ for different tip velocities at 670~K.}
\label{fig:md670}
\end{figure}

\begin{figure}[h!]
\centering
\includegraphics[width=0.47\linewidth]{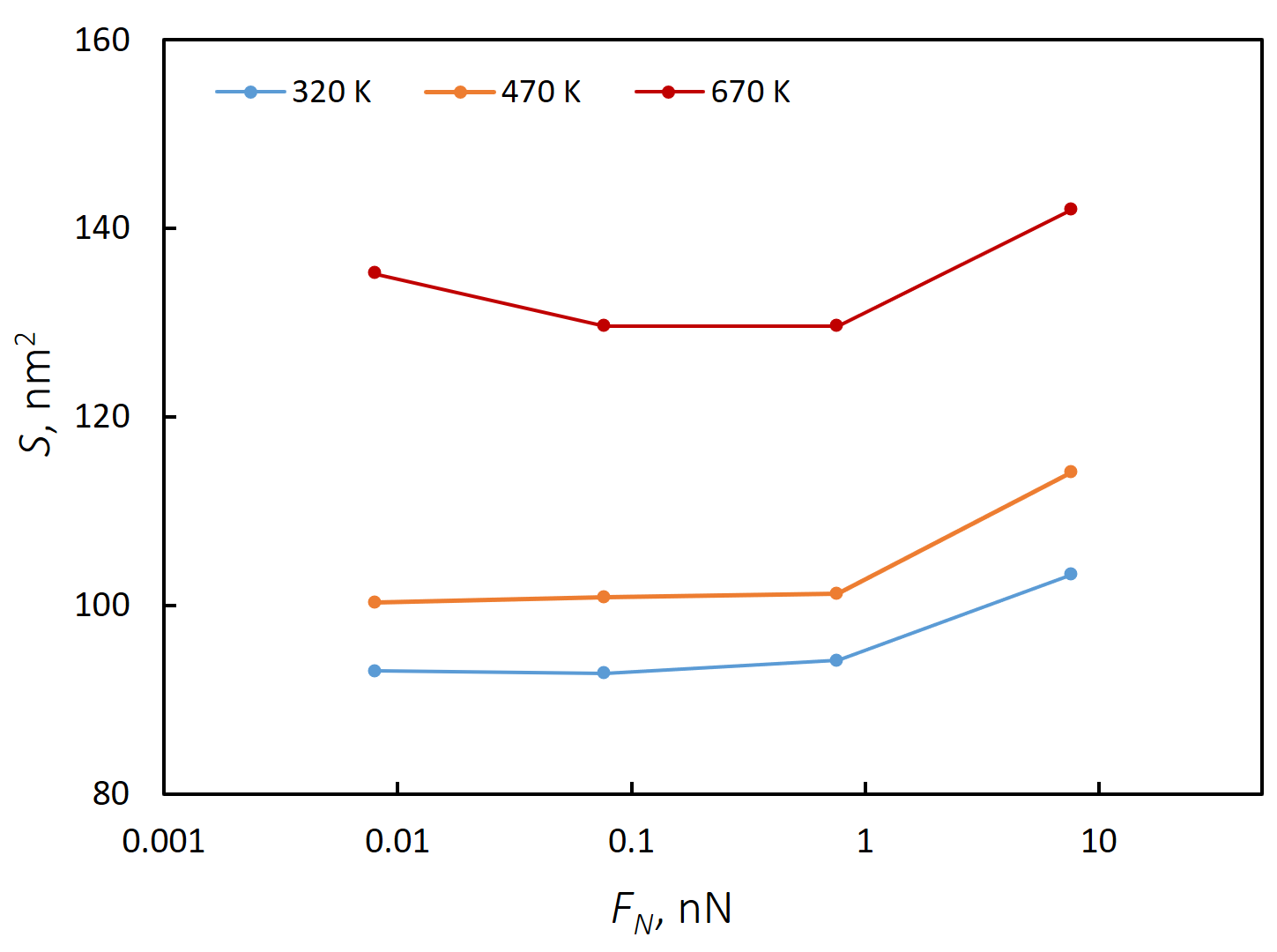} 
\includegraphics[width=0.5\linewidth]{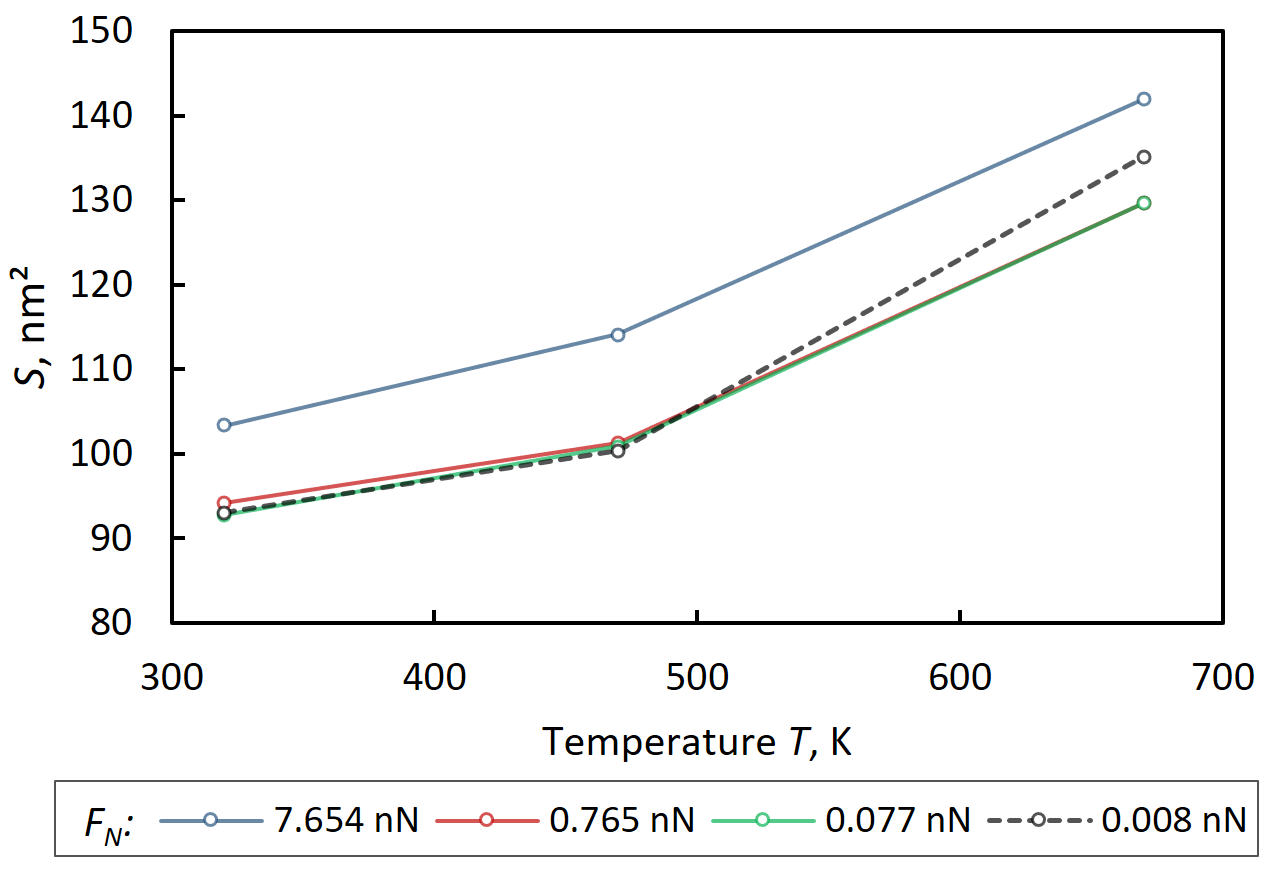}
\caption{\textbf{Contact area as a function of normal load and temperature.} Left panel: The contact area $S$ as a function of the normal load for different temperatures. Note that $S$ only slightly depends on the normal load. Right panel: The contact area as a function of temperature for different normal loads. Note, that the contact area noticeably depends on temperature, which is related to the decrease of material stiffness with increasing temperature.}
\label{fig:ContFn}
\end{figure}

Fig. \ref{fig:md670} shows the dependency of the friction stress $F_{\rm f}/S$ on the  tip's velocity $V_x$  measured for different normal load $F_{\rm N}$ at relatively high temperature of $T \simeq 670$~K, as well as the curves for the friction stress, $F_{\rm f}/S$, versus the normal load $F_{\rm N}$ for different tip velocities (at 670~K).
While the contact area only slightly depends on the normal load, see Fig. \ref{fig:ContFn}, Left panel, it noticeably depends on temperature, Fig. \ref{fig:ContFn}, Right panel. 


\subsection{Friction coefficient as a function of temperature}

\begin{figure}[h!]
\centering
\includegraphics[width=1.0\linewidth]{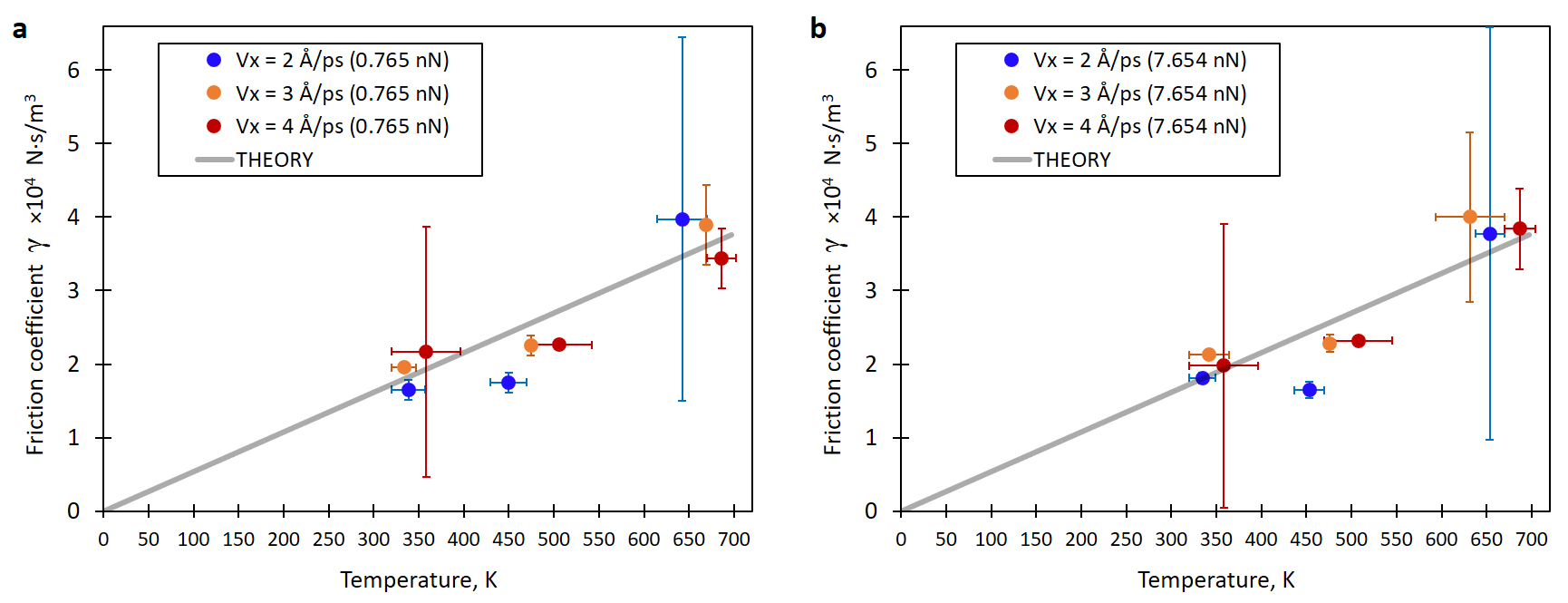}
\caption{\textbf{Friction coefficient as a function of temperature $\gamma = \gamma(T)$.} The estimates of the friction coefficient $\gamma$ from MD simulations for different sliding velocities $V_x$ (shown by markers), and the theoretical linear dependence (shown by grey solid line) for the case of normal load $F_N = 0.765$~nN ({\bf a}) and $F_N = 7.654$~nN ({\bf b}).}
\label{fig:gamma_T}
\end{figure}

The dependence of the friction coefficient, $\gamma=F_{\rm f}/(S V_x)$, on temperature is depicted in Fig.~\ref{fig:gamma_T}. The estimates of the friction coefficient $\gamma$ from MD simulations for different sliding velocities ($V_x=$~2, 3 and 4~{\AA}/ps) are given for the normal load of $F_N = 0.765$~nN and $F_N = 7.654$~nN. The theoretical dependence is shown as a linear function with the same slope,  $\gamma/T = 5.4\cdot 10^{-3}$~$[N\, s/(K \, m^3)]$.

\subsection{Correlation coefficient for the synchronic fluctuations} 
The correlation coefficient $C$, which quantifies the degree of fluctuation synchronicity was evaluated as follows: 
$$
C\equiv\frac{\left(\delta_{\rm I}(x,y,t)\cdot\delta_{\rm II}(x,y,t)\right)}{\sqrt{\left(\delta_{\rm I}(x,y,t)\cdot\delta_{\rm I}(x,y,t)\right)\left(\delta_{\rm II}(x,y,t)\cdot\delta_{\rm II}(x,y,t)\right)}}, 
$$
where $\delta_{\rm I}(x,y,t)$ and $\delta_{\rm II}(x,y,t)$ are respectively the deviations from the equilibrium plane $z=0$ of the central element of the bottom (I) and top (II) surfaces. The scalar product in the above equation is defined as $\left(f_1(x,y,t)\cdot f_2(x,y,t)\right)\equiv\int\left(f_1(x,y,t)f_2(x,y,t)\right){\rm d}t$, where integration  is performed over all simulation time. In practice, we have measured the according deviations of the centers of atoms of the bottom and upper layer from their   equilibrium average  positions; these are shifted, respectively, downwards and upwards, by the atomic radius, from the plane $z=0$. 

We  evaluate $\delta_{\rm I}(x,y,t)$ and $\delta_{\rm II}(x,y,t)$ for two small surface elements $S_c$ of the tip and substrate. We used $z$-positions of several atoms, which belong to the  square region $S_c$ with the dimensions of $6\times 6$~{\AA} in the center of the contact zone. That is, we measured $\delta_{\rm I}(x,y,t)$ as 
$$
\delta_{\rm I}(x,y,t)=\frac{1}{N_{S_c}}\sum_{j\in S_c}\left(z_j^{(I)}(t)-\overline{z_j}^{(I)}\right),
$$
where $N_{S_c}$ is number of atoms in the region $S_c$, and  $ \overline{z_j}^{(I)}$ denotes the is time-average position of atom $j$, belonging to the region $S_c$. Similar definition is applied for $\delta_{\rm I}(x,y,t)$.

\subsection{Substrate with adhered graphene layer as a  joint solid body}

\begin{figure}[h!]
\centering
\includegraphics[width=0.67\linewidth]{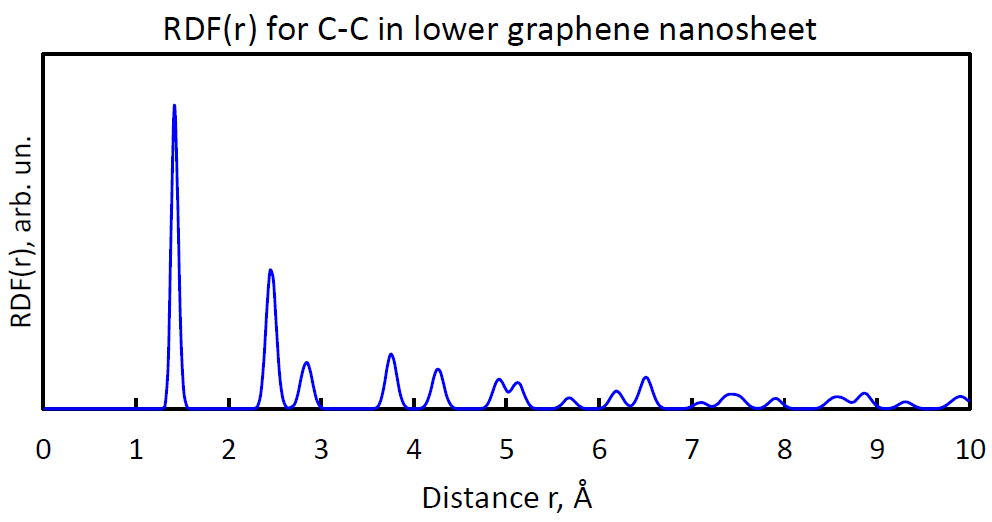}
\caption{\textbf{Radial distribution function for C-C atoms of the lower graphene layer.} The results of the MD experiment for  $T=700$~K, $F_{\rm N} =7.654$~nN and $V_x = 5$~{\AA}/ps are shown.}
\label{fig:rdf-c-c}
\end{figure}

\begin{figure}[h!]
\centering
\includegraphics[width=0.67\linewidth]{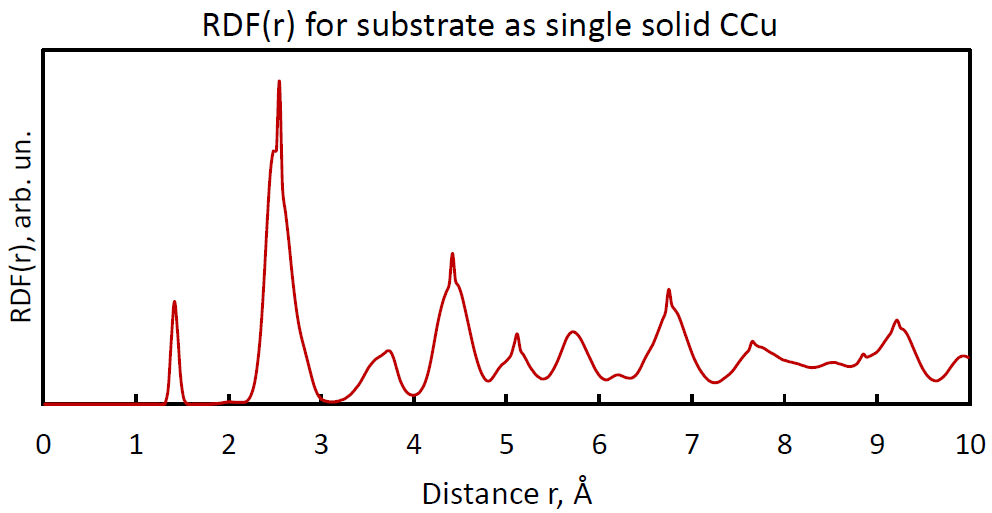}
\caption{\textbf{Radial distribution function for C-Cu -- the carbon atoms of the graphene layer and copper atoms of the substrate.} The RDF shows that the graphene layer and the substrate form a joint solid body.  The results of the MD experiment for  $T=700$~K, $F_{\rm N}=7.654$~nN and $V_x = 5$~{\AA}/ps are shown.}
\label{fig:rdf-ccu}
\end{figure}

\begin{figure}[h!]
\centering
\includegraphics[width=0.67\linewidth]{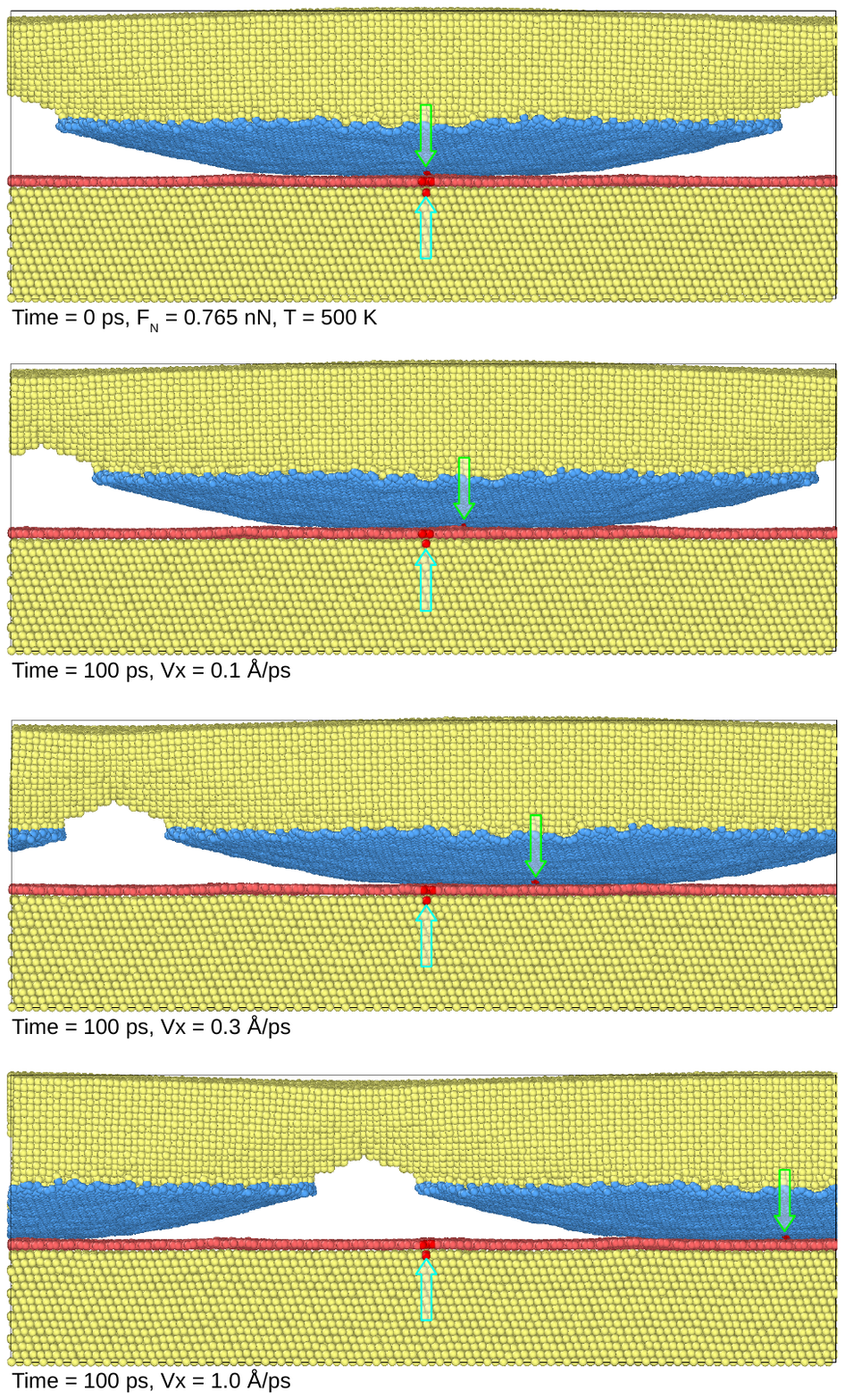}
\caption{\textbf{The analysis of the relative position of labeled carbon and copper atoms for different sliding velocities.} The upper image shows the initial ($t=0$) positions of the labeled groups of atoms on the tip and substrate. Others images depict the system at the time instant $t=100$~ps for the sliding velocities $V_x$ equal   (from top to bottom), to 0.1, 0.3 and 1.0~{\AA}/ps. The light-blue arrow demonstrates that the labeled groups of atoms of the carbon layer and copper atoms of the substrate, do not move with respect to to each other. The light-green arrow indicates the position of the group of labeled carbon atoms, of the graphene layer covering the tip, at the time instant $t=100$~ps. The results of the MD experiment for  $T=500$~K, $F_{\rm N}=0.765$~nN are shown.
}
\label{fig:no-sliding}
\end{figure}

We perform a special analysis to confirm that the  graphene sheet, adhered at the copper single crystal substrate, behave together with the substrate as a joint solid body. Firstly, we measure the  radial distribution function (RDF) C-C within the graphene layer and demonstrate that for  the entire range of the explored  conditions (temperature, normal stress, displacement velocity), it possesses a structure of a two-dimensional crystal, and not of a structureless liquid film, see Fig. \ref{fig:rdf-c-c}. In this figure the RDF is shown for $T = 700$~K, $ F_{\rm N}=7.654$~nN and $V_x = 5$~{\AA}/ps, which corresponds to the largest temperature, normal force and velocity in our simulations -- the least favorable conditions to keep a crystalline structure of the layer. Secondly, we analyze the  RDF for atoms C-Cu -- the carbon atoms of graphene layer and copper atoms of the substrate and again observe a common solid structure, see Fig. \ref{fig:rdf-ccu}; this indicats that the graphene layer and copper substrate, indeed, behave as a joint solid body. Finally, as an additional check, we demonstrate that there is no  sliding of the graphene layer over the copper surface. This has been done, analysing the mutual arrangement of carbon and copper atoms, that is, the relative  position of labeled groups of atoms, see  Figure~\ref{fig:no-sliding}.

\section{Theory} 
In the next Sec. \ref{genthe} we present the derivation of the friction force for the general model. For convenience, in Sec. \ref{simthe} we also give a short derivation of $\langle W_{\rm diss} \rangle$,  given by Eq. 8 of the main text, for the idealized simplified model. 

\subsection{Friction force due to synchronic thermal fluctuations. Analysis of the general case}
\label{genthe}
While in the main text we analyze the idealized model, here we address a general case. We will assume that  the contact of the two surfaces is large enough to obey the laws of continuum elastic theory, which is justified as the continuum theory is already applicable for nanoclusters of about 500 atoms \cite{PRLNano2010}.  
We start from the system, composed of two identical bodies and then explore different bodies. Let ${\bf u}^{(i)}({\bf r},t)$ be the displacement vector of the $i$th  body, $i=1,2$ of the point with the coordinate ${\bf r} =(x,y,z)$ at time $t$. It yields the deformation tensor 
$$
u_{ab} =\frac12 \left(\frac{\partial u_a}{\partial r_b} + \frac{\partial u_b}{\partial r_a} \right),
$$ 
where $a,b=x,y,z$ and $r_a$ are accordingly the $x,y,z$ components of the radius-vector ${\bf r}$. We assume that two bodies are pressed towards each other by the normal load $F_N$ acting along $z$-axis, which yields stress in the bulk of the first and second body with the components $\sigma_{ab}^{(1)}({\bf r})$ and $\sigma_{ab}^{(2)}({\bf r})$.  The deformation energy of two identical bodies at a contact reads, 
\begin{equation}
    \label{E} 
E=  2 \, \frac12 \int_{\Omega} u_{ab}({\bf r}) \sigma_{ab}({\bf r})  d{\bf r}.
\end{equation}
where integration is performed over the volume $\Omega$ of one of the identical bodies and the summation over repeating indexes is implied. The dominant component of the stress tensor for the system of interest is  $\sigma^{(1/2)}_{zz}$; all other components  may be be neglected, as they are significantly smaller than the major one. In this case $u^{\rm eq}_{zz}= \sigma^{\rm eq}_{zz}/Y$, where $Y$ is the Young modulus \cite{Landau:1965}. Hence in the absence of surface thermal fluctuations the two bodies are in a contact on the plane $z=0$ and 
the deformation energy takes the form, 
\begin{equation}
    \label{Eeq} 
E_{\rm eq}= Y \int_{\Omega}  u^{\rm eq}_{zz}({\bf r}) \sigma^{\rm eq}_{zz}({\bf r})  d{\bf r}.
\end{equation}
Suppose the surface fluctuation of the i-the body, $\delta u_z^{(i)}(x,y)$ emerges. It is equal to the deviation of the surface from the equilibrium plane $z=0$ at the point $(x,y)$ and gives rise to the surface corrugation.  The appearance of the fluctuation $\delta u_z^{(i)}(x,y)$ may be  attributed to the fluctuation of the normal    pressure $ \delta P_z^{(i)}(x,y)$, distributed over the contact area as \cite{Landau:1965}, 

\begin{equation}
    \label{Bous} \delta u_z^{(i)}(x,y) =\frac{(1-\nu^2)}{\pi Y} \int_{S} \frac{\delta P_z^{(i)}(x_1,y_1)}{\sqrt{(x-x_1)^2 + (y-y_1)^2}} = \int_S \delta P_z^{(i)}(x_1,y_1) \left. G(x-x_1, y-y_1, z)\right|_{z=0}= \hat{H} \left[ \delta P_z^{(i)} \right], 
\end{equation}
where $\nu$ is the Poisson ratio and integration is performed over the area of  the contact  surface $S$. Here we also introduce the Green function  corresponding to the Boussinesq problem \cite{Landau:1965}:

\begin{equation}
    \label{Green}  G(x, y, z) = \frac{(1+\nu)}{2 \pi Y} \left[ \frac{2(1-\nu)}{(x^2+y^2+z^2)^{1/2}} +\frac{z^2}{(x^2+y^2+z^2)^{3/2}} \right], 
\end{equation}
and the according linear operator $\hat{H}$, corresponding to the convolution of some function and the Green function. Then surface pressure fluctuation may be written as 
\begin{equation}
    \label{inv}   \delta P_z^{(i)}(x,y) = \hat{H}^{-1} \left[ \delta u_z^{(i)} \right],
\end{equation}
where $\hat{H}^{-1}$ is the operator, inverse to $\hat{H}^{-1}$, that is, $\hat{H}\cdot \hat{H}^{-1}=I$. Since the above Green function corresponds to the two-dimensional Coulomb interactions, such an operator exists. Indeed, in the Fourier components Eq. \eqref{Bous} may be written as 
$$
\delta u_z^{(i)} ({\bf q}) =\frac{(1-\nu^2)}{\pi Y} \, \frac{2\pi}{q} \delta P_z^{(i)} ({\bf q}), 
$$
where  ${\bf q}=(q_x, q_y)$, hence $\hat{H}({\bf q}) =2(1-\nu^2) /(Y q) $, and $\hat{H}^{-1}({\bf q}) =(Y q)/( 2(1-\nu^2)) $. 

Using Eq. \eqref{Bous} we can find all components of the  displacement $ \delta {\bf u}^{(i)}({\bf r})$ in the bulk. For instance the component $ \delta u_z^{(i)}({\bf r})$ reads \cite{Landau:1965}:
\begin{equation}
    \label{uzbulk}   
     \delta u_z^{(i)}(x,y,z) = \int_S  \delta P_z^{(i)} (x_1,y_1)  G (x-x_1,y-y_1,z)dx_1dy_1 =  \int_S  \hat{H}^{-1} \left[ \delta u_z^{(i)}\right] (x_1,y_1)  G (x-x_1,y-y_1,z)dx_1dy_1,  
\end{equation}
which expresses the displacement component in the bulk as the function of surface fluctuation at $z=0$. Using Eqs. \eqref{uzbulk}    one can find the components $u_{zz}^{(i)}$, $u_{zx}^{(i)}$ and  $u_{zy}^{(i)}$ of the deformation tensor corresponding to the surface corrugation  $\delta u_z^{(i)}(x,y)$:

\begin{eqnarray}
 \label{deluab}  \delta u_{za}^{(i)} (x,y,z)= \int_S  \delta P_z^{(i)} (x_1,y_1)  \frac{\partial G (x-x_1,y-y_1,z)}{\partial r_a} dx_1dy_1 =  \int_S  \hat{H}^{-1} \left[ \delta u_z^{(i)}\right] (x_1,y_1) \frac{\partial G (x-x_1,y-y_1,z)}{\partial r_a} dx_1dy_1, 
\end{eqnarray}
for $a=x,y,z$ and  $r_a=x,y,z$. The respective components of the stress tensor $\sigma_{ab}(x,y,z)$ read \cite{Landau:1965}, 

\begin{equation}
    \label{sig} \delta \sigma_{zz}^{(i)} = \frac{Y(1-\nu)}{(1+\nu)(1-2 \nu)}\delta u_{zz}^{(i)}; \qquad \delta \sigma_{zx}^{(i)} = \frac{Y}{(1+\nu)}\delta u_{zx}^{(i)}; \qquad \sigma_{zy}^{(i)} = \frac{Y}{(1+\nu)}\delta u_{zy}^{(i)},
    \end{equation}
where  $\delta u_{za}^{(i)} (x,y,z)$ with $a=x,y,z$  are given in the above Eq. \eqref{deluab} 

Now we write down the change of the deformation energy of the system owing to the surface fluctuations $\delta u_z^{(1)}(x,y)$ and $\delta u_z^{(2)}(x,y)$. Using Eqs. \eqref{E}, \eqref{Eeq} and \eqref{sig} we obtain, 

\begin{eqnarray}
    \label{delE} \delta E &=&\frac12 \sum_{i=1,2} \sum_{a=x,y,z} \int_{\Omega_i} d{\bf  r}_i  (\sigma_{za}^{\rm eq } +\delta \sigma_{za}^{(i)} )( u_{za}^{\rm eq} +\delta u_{za}^{(i)})  - E_{\rm eq}\\
    &=& \frac12 \left[\int_{\Omega_1} \sigma_{zz}^{\rm eq } \delta u_{zz}^{(1)} d{\bf r}_1 + \int_{\Omega_2} \sigma_{zz}^{\rm eq } \delta u_{zz}^{(2)} d{\bf r}_2 \right] 
    + \frac{2 Y(1-\nu)}{2(1+\nu)(1-2 \nu)} \left[\int_{\Omega_1} u_{zz}^{\rm eq } \delta u_{zz}^{(1)} d{\bf r}_1 + \int_{\Omega_2} u_{zz}^{\rm eq } \delta u_{zz}^{(2)} d{\bf r}_2 \right] \nonumber \\
    &+&\frac{2 Y(1-\nu)}{2(1+\nu)(1-2 \nu)} \left[\int_{\Omega_1} \left( \delta u_{zz}^{(1)}\right)^2  d{\bf r}_1 + \int_{\Omega_2} \left( \delta u_{zz}^{(2)} \right)^2 d{\bf r}_2 \right] \nonumber \\
    &+&\frac{Y}{2(1+\nu)} \left[\int_{\Omega_1} \left( \delta u_{zx}^{(1)}\right)^2  d{\bf r}_1 + \int_{\Omega_2} \left( \delta u_{zx}^{(2)} \right)^2 d{\bf r}_2 \right] + 
    \frac{Y}{2(1+\nu)} \left[\int_{\Omega_1} \left( \delta u_{zy}^{(1)}\right)^2  d{\bf r}_1 + \int_{\Omega_2} \left( \delta u_{zy}^{(2)} \right)^2 d{\bf r}_2 \right]  \nonumber 
       \end{eqnarray}
where $\sigma_{za}^{\rm eq }= \sigma_{zz}^{\rm eq }\delta_{za}$ and $u_{za}^{\rm eq}=u_{zz}^{\rm eq}\delta_{za}$. 

Consider now synchronic fluctuations, such that $\delta u_z^{(1)} (x,y) )=-\delta u_z^{(2)} (x,y)$. Due to the linearity of Eqs. \eqref{inv} and \eqref{deluab}, we conclude that $\delta u_{zz}^{(1)}=-\delta u_{zz}^{(2)}$, $\delta u_{zx}^{(1)}=-\delta u_{zx}^{(2)}$ and $\delta u_{zy}^{(1)}=-\delta u_{zy}^{(2)}$. As the result, the linear order terms with respect to $\delta u_{zx}^{(1/2)}$ vanish in the above equation, which means that the energy of  the synchronic surface fluctuation does not have linear in the fluctuations terms.  That is, it reads, 
\begin{equation}
    \label{delE2} \delta E = \frac{Y(1-\nu)}{(1+\nu)(1-2 \nu)} \int_{\Omega_1} \left( \delta u_{zz}^{(1)}\right)^2  d{\bf r}_1 + \frac{ Y}{(1+\nu)} \int_{\Omega_1} \left( \delta u_{zy}^{(1)}\right)^2  d{\bf r}_1 +\frac{ Y}{(1+\nu)} \int_{\Omega_1} \left( \delta u_{zx}^{(1)}\right)^2  d{\bf r}_1, 
\end{equation}
The important consequence of Eq. \eqref{delE2} is that the energy of synchronic surface fluctuations {\it does not} depend on the equilibrium stress $\sigma_{zz}^{\rm eq}$, that is, on the normal load $F_N$.  At the same time, the values of  $ \delta u_{zz}^{(1)}$,  $ \delta u_{zy}^{(1)}$  and $\delta u_{zx}^{(1)}$ is this equation are determined by the probability of thermal fluctuations as  discussed below. Note that in the above equation we do not account for the variation of the surface area due to the fluctuations and the respective increase of the surface energy. This is justified due to the tight contact of the two surfaces, which would be not true if the fluctuations were not synchronic.  

Now we consider the relative motion of two surfaces with the velocity $V_x=V>V_*$, along the axis $x$, where $V_*$ refers to the propagation speed of the synchronic surface fluctuations. For $V_x>V_*$ the fluctuation do not have time to relax hence the surface profile reads, 
$$
u_z(x,y,t)= \delta u_z(x-Vt, y), 
$$
that is, the relative motion drives the surface thermal "wrinkles" retaining their shape. Hence,  the running surface fluctuation  $\delta u_z(x-Vt, y)$ may be attributed to the running pressure fluctuation $P_z^{(i)}(x_1-Vt,y_1)$.  Correspondingly, Eq. \eqref{uzbulk} yields for the  bulk component of the displacement, $u_z^{(i)}(x,y,z)$: 
\begin{equation}
    \label{uzdt} \delta u_z^{(i)}(x,y,z)= \int_S \delta P_z^{(i)}(x_1-Vt,y_1) G(x-x_1, y-y_1, z)dx_1dy_1. 
\end{equation}
For the time derivative of this quantity we obtain,  
\begin{eqnarray}
 \label{derdel}  \frac{\partial }{\partial t} \delta u_{z}^{(i)} (x,y,z) &=& -V  \int_S  G(x-x_1, y-y_1, z) \frac{\partial \delta P_z^{(i)}(x_1-Vt,y_1)}{\partial x_1} dx_1dy_1 \\ 
  &=& 
 V \int_S  \delta P_z^{(i)}(x_1-Vt,y_1)  \frac{\partial G(x-x_1, y-y_1, z)}{\partial x_1}  dx_1dy_1 \nonumber  \\ 
 &=& - V \frac{\partial }{\partial x} \int_S \delta P_z^{(i)}(x_1-Vt,y_1) G(x-x_1, y-y_1, z)dx_1dy_1= -V  \frac{\partial }{\partial x} \delta u_{z}^{(i)} (x,y,z), \nonumber
\end{eqnarray}
where we apply the integration by parts and use the condition $\delta P_z^{(i)}=0$ on the boundary of the contact domain. Taking into account that the time and space derivatives commute, we conclude that the time derivative of the deformation reads, 
\begin{equation}
    \label{dotu} \delta \dot{u}_{zz}^{(i)} = -V \frac{\partial }{\partial x} \delta u_{zz}^{(i)};  \qquad \delta \dot{u}_{zx}^{(i)} = -V \frac{\partial }{\partial x} \delta u_{zx}^{(i)}; \qquad \delta \dot{u}_{zy}^{(i)} = -V \frac{\partial }{\partial x} \delta u_{zy}^{(i)},  
\end{equation}
where the upper dot denotes the time derivative.

Due to the relative motion  of the two surfaces the deformation in the both contacting bodies varies with time. This causes the dissipation of the mechanical energy, quantified by the viscosity of the bodies material (the same process is associated with the dissipation of sound in solids). The dissipation power reads, 
\begin{equation}
    \label{Wdiss}
W_{\rm dis} = \int_{\Omega}   R({\bf r})  d{\bf r}  ; \qquad \qquad R= \eta_1 \left( \dot{u}_{ab} -\tfrac{1}{3} \delta_{ab} \dot{u}_{cc} \right)^2 + \tfrac12 \eta_2 \dot{u}_{cc}^2,
\end{equation}
where $\eta_1$ and $\eta_2$ are coefficients of solid viscosity, $\delta_{ab}$ is the Kronecker symbol, $a,b,c =x,y,z$ and Einstein summation convention is applied. In our case the non-zero components of the deformation rate are $\delta \dot{u}_{zx}^{(i)}$,  $\delta \dot{u}_{zy}^{(i)}$ and  $\delta \dot{u}_{zz}^{(i)}$ for $i=1,2$. Then we obtain for the identical bodies, 
\begin{equation}
    \label{Wdiss1}
W_{\rm dis} = 2 \int_{\Omega}     \left[ \eta_1 \left(  \delta \dot{u}_{zx} \right)^2 + \eta_1 \left(  \delta \dot{u}_{zy} \right)^2 + \eta \left(  \delta \dot{u}_{zz} \right)^2 \right] d{\bf r} 
=2 V^2 \int_{\Omega} \left[ \eta_1 \left(  \frac{\partial }{\partial x}  \delta {u}_{zx} \right)^2 + \eta_1 \left( \frac{\partial }{\partial x}   \delta {u}_{zy} \right)^2 + \eta \left(  \frac{\partial }{\partial x} \delta {u}_{zz} \right)^2 \right] d{\bf r} ,
\end{equation}
where $\eta \equiv  (\tfrac49 \eta_1 + \tfrac12 \eta_2)$, we skip the index $i$ indicating the identical bodies, and use Eq. \eqref{dotu}. 

Eq. \eqref{Wdiss1} demonstrates that the dissipation depends on the relative velocity $V$ and the surface fluctuation profile $\delta u_z(x,y)$, which determines the bulk deformation $\delta u_za(x,y,z)$, with $a=x,y,z$ and the according deformation rates. The quantity $\delta u_z(x,y)$ is random, as it is caused by the thermal noise. Hence, one needs to average over all possible values of $\delta u_z(x,y)$ with the respective probability density. The probability of $\delta u_z(x,y)$ is determined by the energy $\delta E$, associated with this fluctuation, given by Eq. \eqref{delE2}; it  follows from the basic statistical mechanics \cite{Feynman}: 
\begin{equation}
    \label{Prob} P[\delta u_z(x,y)] =\frac{1}{Z} e^{-\delta E[\delta u_z(x,y)]/k_BT } \qquad \qquad Z= \int {\cal D} [\delta u_z(x,y)] e^{-\delta E[\delta u_z(x,y)]/k_BT }, 
\end{equation}
where $k_B$ is the Boltzmann constant, $T$ is temperature, and ${\cal D} [\delta u_z(x,y)]$ -- designates the integration over functions. That is, that the integration is to be performed over all possible two-dimensional functions, defined on the contact area $S$, which are associated with  the surface fluctuations. Integration over the functions may be performed using the standard techniques, see e.g.  \cite{Feynman}. The average dissipation power then reads, 
\begin{equation}
    \label{Wdiss2} \left\langle W_{\rm diss} \right \rangle = F_f V =     \int  {\cal D} [\delta u_z(x,y)] \,  W_{\rm diss} [\delta u_z(x,y)]\,\ \, P[\delta u_z(x,y)] ,  
\end{equation}
where we take into account that the dissipation power is equal to the product of the  dissipative force, here -- the friction force, and the velocity. Using Eqs. \eqref{Wdiss1}, \eqref{Prob} and \eqref{Wdiss2}, we can obtain, 
\begin{equation}
    \label{FgV} F_f=\gamma S V, \qquad \qquad 
    \gamma=\frac{2}{ZS} \int  {\cal D} [\delta u_z]    \left\{ \int_{\Omega} \left[ \eta_1 \left(  \frac{\partial }{\partial x}  \delta {u}_{zx} \right)^2 + \eta_1 \left( \frac{\partial }{\partial x}   \delta {u}_{zy} \right)^2 + \eta \left(  \frac{\partial }{\partial x} \delta {u}_{zz} \right)^2 \right] d{\bf r} \right\} e^{-\delta E[\delta u_z]/k_BT }.  
\end{equation}
Note that in the above expression for the friction force $F_f$ there are no terms that depend on the normal load $F_N$. Hence, Eq. \eqref{FgV} proves the independence of the friction force on the load for the dominating synchronic fluctuations. Moreover, it shows linear dependence on the velocity $V$, if $V>V_*$. The linear dependence of $F_f$ on the area of the surface contact $S$ will follow from the independence of $\gamma$ on $S$. It is still to be proved by the direct calculation of $\gamma$.  In the main text has been computed for the case of uniform deformation in the both contacting bodies. 

Let us perform the respective computation of $\gamma$. The standard way to integrate over functions is to apply the Fourier transform to the functions and then perform the conventional integration over the according Fourier coefficients, see e.g. \cite{Feynman}.  We will assume that the contact area is large enough, so that the standard technique is applicable and will represent the dependence on the coordinates $x$ and $y$, specifying the contact in terms of the plane waves $ e^{iq_xx+iq_yy}/S= e^{i {\bf q} \cdot  {\bf \rho}}/S$, with the discrete wave vectors ${\bf q}$; here ${\bf \rho}=(x,y)$ is the two-dimensional vector, such that ${\bf r}=(x,y,z)=({\bf \rho}, z)$ and $S$ the area of the contact. Then the Fourier transform of Eq. \eqref{Bous} reads 
\begin{equation}
    \label{FTu} \delta u_z({\bf q}) =\frac{(1-\nu^2)}{\pi Y} \delta P_z({\bf q}) G({\bf q},z=0), 
\end{equation}
where we skip for brevity the body index $i$ and use the theorem for the Fourier transform for the convolution of two functions. The Fourier transform of the Green function is: 
\begin{eqnarray}
    \label{FTG} 
 G({\bf q},z)&= & \int_S e^{-{\bf q} \cdot {\bf \rho}} G(x,y,z) \approx \int_0^{2 \pi} d \varphi \int_0^{\infty} d\rho \, \rho e^{-i q \rho \cos \varphi}  \left[\frac{a_1 }{(\rho^2+z^2)^{1/2}}+\frac{a_2 \, z^2}{(\rho^2+z^2)^{3/2}}  \right] \\ &= &2 \pi \int_0^{\infty} d\rho  J_0(q\rho)\,  \left[\frac{a_1 \rho }{(\rho^2+z^2)^{1/2}}+\frac{a_2 \rho\, z^2}{(\rho^2+z^2)^{3/2}}  \right]
 = 2 \pi e^{-qz} \left[\frac{a_1}{q} + a_2z \right] \nonumber , 
\end{eqnarray}
where $J_0(x)$ is the Bessel function. Here we again assume that the contact area is large, so that one can approximate the Fourier  transform for a finite area by the Fourier transform for a circular domain with the radius that tends to infinity radius; we also  abbreviate 
$$
a_1= (1-\nu^2)/(\pi Y) \qquad \qquad a_2= (1+\nu)/(2 \pi Y).
$$ 
Hence $G({\bf q},z=0)= 2\pi/q$ and we find from the above equations: 
\begin{equation}
    \label{dPzq} 
     \delta P_z({\bf q}) = \frac{Yq}{2(1-\nu^2)}  \delta u_z({\bf q}) = \frac{q}{2\pi a_1} \delta u_z({\bf q}) . 
\end{equation}
To find the Fourier transform of $\delta u_{za} (x,y,z)$ for $a=x,y,z$, we need to use the Fourier transform of the according derivatives of $G(x,y,z)$, as it follows from Eq. \eqref{deluab}. Performing the same computations as in Eq. \eqref{FTG}, we find: 
\begin{equation}
    \label{FTGder} 
 G_x({\bf q},z) = 2 \pi i [a_1+a_2qz]e^{-qz}; \qquad \qquad G_y({\bf q},z)  =0; \qquad \qquad  G_z({\bf q},z) = 2 \pi  [a_2(1-qz) -a_1]e^{-qz},  \end{equation}
where $G_a= \partial G/\partial r_a$.  

Using the Fourier transform of Eq. \eqref{deluab} which contains the convolution of two functions, 
$$
\delta u_{za}({\bf q},z) = \frac{q}{2 \pi a_1} \delta u_z({\bf q}) G_a ({\bf q},z), 
$$
we obtain
\begin{equation}
    \label{uzzr} 
\delta u_{zz}(x,y,z)= \frac{1}{S} \sum_{\bf q} e^{iq_x x+ iq_y y }   \frac{q}{2 \pi a_1} \delta u_z({\bf q}) G_z ({\bf q},z)=\frac{1}{S} \sum_{\bf q} e^{i{\bf q} \cdot \bf{\rho}}   \frac{q}{2 \pi a_1} \delta u_z({\bf q}) G_z ({\bf q},z).
\end{equation}
Hence, 
\begin{equation}
    \label{uzzint} 
\int_{\Omega} d {\bf r} \left( \delta u_{zz}({\bf r} )\right)^2 = \frac{1}{S^2} \sum_{\bf q}  \sum_{{\bf q}_1} \int dz \int d{\bf \rho} 
\frac{qq_1}{(2 \pi a_1)^2 } G_z ({\bf q},z) G_z ({\bf q}_1 ,z) \delta u_z({\bf q}) \delta u_z({\bf q}_1) e^{i  \bf{\rho} \cdot  ({\bf q} +{\bf q}_1) }. 
\end{equation}
With the relation
$
\int d{\bf \rho } e^{i  \bf{\rho} \cdot  ({\bf q} +{\bf q}_1) } = S\delta_{{\bf q}, -{\bf q}_1}, 
$
where $\delta_{{\bf a},{\bf b}}$ is the Kronecker delta and Eq. \eqref{FTGder}, we obtain, 
\begin{equation}
    \label{uzzint1} 
\int_{\Omega} d {\bf r} \left( \delta u_{zz}({\bf r} )\right)^2 = 
\frac{1}{S} \sum_{\bf q} \frac{q^2}{(2 \pi a_1)^2 } [\delta u_z({\bf q})]^2 \int_0^{\infty}  \left[2\pi(a_2(1-qz)+a_1e^{-qz} \right]^2 dz = \frac{1}{4S} \sum_{\bf q} \frac{(5-8\nu +8 \nu^2)}{4(1-\nu)^2} q [\delta u_z({\bf q})]^2 .
\end{equation}
Similarly, we find 
\begin{equation}
    \label{uzxint1} 
\int_{\Omega} d {\bf r} \left( \delta u_{zx}({\bf r} )\right)^2 = \frac{1}{4S} \sum_{\bf q} \frac{(13-20\nu +8 \nu^2)}{4(1-\nu)^2} q [\delta u_z({\bf q})]^2 .
\end{equation}
Therefore, with $ \delta u_{zy}({\bf r}) =0$ [recall that $G_y({\bf q},z)=0$, see Eq. \eqref{FTGder}] we find, 
\begin{eqnarray}
    \label{delE1} 
\delta E[\delta u_z({\bf q})]&=& \frac{Y(1-\nu)}{(1+\nu) (1-2 \nu)} \frac{1}{4S} \sum_{\bf q} \frac{(5-8\nu +8 \nu^2)}{4(1-\nu)^2} q [\delta u_z({\bf q})]^2 + \frac{Y}{(1+\nu)}  \frac{1}{4S} \sum_{\bf q} \frac{(13-20\nu +8 \nu^2)}{4(1-\nu)^2} q [\delta u_z({\bf q})]^2 \\ &=&
\frac{BY}{S} \sum_{\bf q}  q [\delta u_z({\bf q})]^2 \nonumber,
\end{eqnarray}
where 
$$
B=B(\nu)= \frac{1}{4(1+\nu)(1-\nu)^2}\left[(13 -20 \nu +8\nu^2) +\frac{(1-\nu)}{(1-2\nu)}(5-8\nu+8\nu^2) \right].
$$
Note that this result has been obtained for vector ${\bf q}$ chosen in the direction of axis $x$. It is easy to show that the result does not depend on the chosen direction of ${\bf q}$. 

Now we compute the square average of the Fourier transform of the deformation $[\delta u_z({\bf q})]^2$. Using the probability distribution \eqref{Prob} and changing from the integration over the functions to the integration over the Fourier components, see e.g. \cite{Feynman}, we obtain: 
\begin{equation}
    \label{uzqsq} 
    \left\langle [\delta u_z({\bf q})]^2 \right\rangle = \frac{1}{Z} \int \prod_{\bf q} d\delta u_z({\bf q})\,  e^{ - \delta E[\delta u_z({\bf q})]}  [\delta u_z({\bf q})]^2  ; \qquad \qquad Z \equiv \int \prod_{\bf q} d\delta u_z({\bf q}) \, e^{ - \delta E[\delta u_z({\bf q})]/k_BT}   ,
\end{equation}
where the Fourier coefficients $\delta u_z({\bf q})$ of the surface thermal fluctuations are the random variables. Changing the integration variables, $\delta u_z({\bf q}) \to y_{\bf q}$  and using Eq. \eqref{delE1}, the integrals in the above equation may be written as 
\begin{equation}
    \label{uzqsq1} 
    \left\langle [\delta u_z({\bf q})]^2 \right\rangle = \frac{1}{Z} \int \prod_{\bf q} d y_{\bf q} \,  e^{ - (BY/k_BT S) \,  q y_{\bf q}^2 }   y_{\bf q}^2    ; \qquad \qquad Z \equiv \int \prod_{\bf q} d y_{\bf q} \,  e^{ - (BY/k_BT S)\,  q y_{\bf q}^2 } .
\end{equation}
These Gaussian integrals are easily computed, yielding 
\begin{equation}
    \label{uzqsq2} 
    \left\langle [\delta u_z({\bf q})]^2 \right\rangle = \frac{k_BT S}{2qBY}.
\end{equation}

Next we need to compute $\partial \delta u_{zz} /\partial x $ and $\partial \delta u_{zx} /\partial x $. From Eq. \eqref{uzzr} it follows that 
\begin{equation}
    \label{duzzdx} 
\frac{ \partial}{ \partial x} \delta u_{zz}(x,y,z)= \frac{1}{S} \sum_{\bf q} i q_x e^{i{\bf q} \cdot \bf{\rho}}   \frac{q}{2 \pi a_1} \delta u_z({\bf q}) G_z ({\bf q},z).
\end{equation}
with the similar result for $\partial \delta u_{zx} /\partial x $. Again, similarly to Eq. \eqref{uzzint}  we write, 
\begin{equation}
    \label{duzzxint} 
\int_{\Omega} d {\bf r} \left( \frac{ \partial}{ \partial x} \delta u_{zz}({\bf r} )\right)^2 = \frac{1}{S^2} \sum_{\bf q}  \sum_{{\bf q}_1} \int dz \int d{\bf \rho} 
\frac{q_x q_{1\, x}\, q q_1}{(2 \pi a_1)^2 } G_z ({\bf q},z) G_z ({\bf q}_1 ,z) \delta u_z({\bf q}) \delta u_z({\bf q}_1) e^{i  \bf{\rho} \cdot  ({\bf q} +{\bf q}_1) }. 
\end{equation}
Performing the same steps which led from Eq. \eqref{uzzint} to Eq. \eqref{FTGder}, we obtain: 
\begin{equation}
    \label{duzzxint1} 
\int_{\Omega} d {\bf r} \left( \frac{ \partial}{ \partial x} \delta u_{zz}({\bf r} )\right)^2 = \frac{1}{4S} \sum_{\bf q} \frac{(5-8\nu +8 \nu^2)}{4(1-\nu)^2} q_x^2q [\delta u_z({\bf q})]^2 .  
\end{equation}
In the same way we find, 
\begin{equation}
    \label{duzxxint1} 
 \int_{\Omega} d {\bf r} \left( \frac{ \partial}{ \partial x} \delta u_{zx}({\bf r} )\right)^2   = \frac{1}{4S} \sum_{\bf q} \frac{(13-20\nu +8 \nu^2)}{4(1-\nu)^2} q_x^2 q [\delta u_z({\bf q})]^2 .
\end{equation}

Using Eq. \eqref{FgV} one can write the friction coefficient, 
\begin{equation}
    \label{gam} 
        \gamma=\frac{2}{S} \left[ \eta_1 \left\langle \int_{\Omega} d {\bf r} \left( \frac{ \partial}{ \partial x} \delta u_{zx}({\bf r} )\right)^2 
        \right \rangle + \eta \left\langle \int_{\Omega} d {\bf r} \left( \frac{ \partial}{ \partial x} \delta u_{zz}({\bf r} )\right)^2 \right \rangle \right] = \frac{\eta }{2S^2} \sum_{\bf q} q_x^2 q B_1 \left\langle  [\delta u_z({\bf q})]^2 \right \rangle, 
\end{equation}
where we define 
\begin{equation}
    \label{B1} B_1\equiv \frac{(5-8\nu +8 \nu^2)}{4(1-\nu)^2} +  \left( \frac{ \eta_1}{\eta} \right)\, \frac{(13-20\nu +8 \nu^2)}{4(1-\nu)^2}, \end{equation}
and use Eqs. \eqref{duzzxint1} and \eqref{duzxxint1}. Applying now Eq. \eqref{uzqsq2} for $\left\langle  [\delta u_z({\bf q})]^2 \right \rangle$ we arrive at, 
\begin{equation}
    \label{gam1} 
        \gamma= \frac{B_1}{B} \frac{ \eta k_BT}{4Y S} \sum_{\bf q} q_x^2. 
\end{equation}
We assume that the area $S$ of the  is large, so that the  summation over the discrete set  of vectors, ${\bf q}=(q_x,q_y)$ may be approximated by the integration over the continuum vector ${\bf q}$. This is justified as for large systems the difference between the discrete vectors ${\bf q}$ is small.  Performing the integration, one should,  however, employ the factor  $S/(4 \pi^2)$, which accounts for the density of discrete vectors in the two dimensional ${\bf q}$-space, see e.g. \cite{SolidSt}. Hence we obtain, 
\begin{equation}
\label{qmax}
\sum_{\bf q} q_x^2 \simeq \frac{S}{4 \pi^2} \int_0^{2\pi} d \phi \cos^2 \phi  \int_0^{q_{\rm max}}q^3 dq = \frac{S}{16 \pi}q^4_{\rm max},
\end{equation}
where  $q_{\rm max}$ is the maximal wave-vector for the system. It refers to the shortest characteristics length, equal to the lattice constant $d$, i.e.   $q_{\rm max} \simeq 2\pi/d$. Alternatively, it may be assessed through the total number of degrees of freedom, associated with the total number of the discrete wave-vectors: 
$$
2N_s= \frac{S}{4\pi^2}\int_0^{2\pi}d \phi \int_0^{q_{\rm max}}q dq = \frac{S}{4\pi} q_{\rm max}.
$$
Here in the left-hand side we have the total number of the surface degrees of freedom, equal to $2N_s$, where $N_s$  is the number of the surface atoms for the area $S$. Hence $q_{\rm max}= 2 \sqrt{2 \pi \rho_s}$, where  $\rho_s= N_s/S$ is the density of surface atoms. From Eqs.\eqref{gam1} and \eqref{qmax} we finally obtain, 
\begin{equation}
    \label{gamfin} 
        \gamma= \pi b \frac{k_BT}{Y}\eta \rho_s^2; \qquad \qquad \qquad 
        b=  \frac{B_1}{B\pi^2}= \frac{(1+\nu)(1-2\nu)}{\pi^2}\,\left[  \frac{b_1+rb_2}{b_2+(1-\nu)b_1 }\right],
\end{equation}
with $b_1\equiv 5-8\nu +8 \nu^2$ and $b_2=13-20\nu +8 \nu^2$. The coefficient $r$ depends on the ratio of the friction coefficients, $r=(\tfrac49 +\tfrac12 \eta_2/\eta_1)^{-1}$.

The appearance of $z$-component of the surface fluctuation implies also the emergence of other components, $\delta u_x^{(i)}(x,y)$ and $\delta u_y^{(i)}(x,y)$. These may be found with the use of the  Boussinesq solution \cite{Landau:1965}. In principle, it is straightforward to perform exactly the same calculations as have been performed for the $z$ component  of the displacement. Their contribution to the friction force are however considerably less significant and hence may be neglected. 

\subsection{Calculation of $\langle W_{\rm diss} \rangle$ for the simplified theory of the main text}
\label{simthe}

We write the surface fluctuations as a sum of cosine and sine functions of $x$ and $y$, that is, we decompose $\delta(x,y)$ in the Fourier series: 
\begin{equation}
\label{del}
\delta(x,y) =\sum_{k_x} \sum_{k_y}  \left[ A_{k_x,k_y }\cos k_x x\cos k_y y +B_{k_x,k_y }\cos k_x x\sin k_y y    +C_{k_x ,k_y }\sin k_x x\cos k_y y   +D_{k_x ,k_y  }\sin k_x x\sin k_y y \right]. 
\end{equation}
Using the relation for $\delta E$ in the main text and the equivalence of all four terms in \eqref{del}, it is straightforward to find the energy of such fluctuations, 
$$
\delta E=  4\,  S(Y/L) \sum_{k_x} \sum_{k_y} \tfrac14 A_{k_x,k_y }^2 .
$$ 
The probability of the set of the amplitudes $\{A_{k_x,k_y}\}$ reads, 
\begin{eqnarray}
\label{Ak} 
          P\left[ \{A_{k_x,k_y}\} \right] = B\,e^{-\delta E/k_BT}  =Z^{-1} e^{-YS/L \sum_{k_x} \sum_{k_y} A_{k_x,k_y }^2/k_BT}; \qquad \qquad 
     Z=  \int \prod_{k_x}\, \prod_{k_y}d A_{k_x,k_y }e^{-YS/L \sum_{k_x} \sum_{k_y} A_{k_x,k_y }^2/k_BT}, \nonumber
\end{eqnarray}
which yields the square average amplitude of the  Fourier mode: 
\begin{equation}
    \langle A^2_{\bf k} \rangle = Z^{-1}\int \prod_{\bf k}  A^2_{\bf k}d A_{\bf k} e^{-(YS/Lk_BT) \sum_{\bf k} A^2_{\bf k}} = \frac{L k_B T}{2 SY}, \end{equation}
where we abbreviate, ${\bf k}=(k_x,k_y)$. Now we can write, 
\begin{equation}
\begin{split}
\label{intdel}
   \left \langle \int \left( \frac{\partial \delta(x,y) }{\partial x } \right)^2 dx dy \right \rangle &=  4  \left \langle \int dx dy \left[ \sum_{k_x}\sum_{k_y}  k_x 
    A _{k_x,k_y} \sin k_x x \cos k_y y \right]^2 \right\rangle  \\ &= 4 S \left \langle \sum_{\bf k} \tfrac14 k_x^2 A^2_{\bf k}  \right \rangle = \frac{L k_B T}{2Y} \frac12 \sum_{\bf k} {\bf k}^2, 
\end{split}
\end{equation}
where ${\bf k}^2=k_x^2+k_y^2$ and we use the above expression for  $\langle A^2_{\bf k} \rangle$; we also exploit the symmetry of fluctuations with respect to the directions $x$ and $y$. Assuming that the area $S$ of the contact surface is large, we approximate the summation over the discrete set  of vectors, ${\bf k}=(k_x,k_y)$ by the integration over the continuum vector ${\bf k}$. This is justified as the difference between discrete vectors ${\bf k}$ is small for large systems, e.g. \cite{SolidSt,BrilFarad}. Hence we obtain, 
\begin{equation}
\label{kmax}
\sum_{\bf k} {\bf k}^2 \simeq \frac{S}{4 \pi^2} \int_0^{2\pi} d \phi  \int_0^{k_{\rm max}}k^3 dk = \frac{S}{8 \pi}k^4_{\rm max},
\end{equation}
where the factor $S/(4 \pi^2)$ accounts for the density of discrete vectors in two dimensional ${\bf k}$-space, when changing from the summation to integration \cite{SolidSt,BrilFarad} and $k_{\rm max}$ is the maximal wave-vector for the system. It refers to the shortest characteristics length, equal to the lattice constant $d$, i.e.   $k_{\rm max} \simeq 2\pi/d$. Alternatively, it may be assessed through the total number of degrees of freedom associated with the total number of discrete wave-vectors, yielding $k_{\rm max} \simeq 2 \sqrt{ 2 \pi \rho_s}$, where  $\rho_s$ is the density of surface atoms. Using Eqs. \eqref{intdel}, \eqref{kmax} and the relation for $W_{\rm diss}$ in the main text, we finally obtain,  
\begin{equation}
\label{Wdiss4}
\langle W_{\rm diss} \rangle  = \left(\frac{2 \eta}{L} \right) \, b \, \frac{L k_B T}{2Y} \frac12 \, \frac{S}{8 \pi} \, 16 \rho_s^2 \pi^2   =\pi \, b\, \frac{k_BT}{Y}\rho_s^2 \eta V^2 S, 
\end{equation}
where $b=1$ for the simplified model. Calculations  of the friction coefficient $\gamma$ for the general model, detailed above, yield $b$ given by Eq. \eqref{gamfin}.

\subsection{Estimates of the threshold velocity $V_*$} 

To estimate the threshold velocity $V_*$ we use the numerically obtained relaxation time,  $\tau$, of the synchronic fluctuations and their characteristic height,  $\delta$. As it may be seen from Fig. 3 of the main text, these quantities may be estimated as $\tau \sim 5-10$ ps and $\delta \sim 0.1 - 0.2$ \AA. We need to estimate the  characteristic length $l$ of the surface thermal excitations. We assume that the continuum elastic theory is applicable and consider a fluctuation (deformation) of the  Boussinesq form. This refers to the  uniform displacement of a circular area of a plane, which is the surface of an elastic semi-space is in the direction, which normal to the plane. For this type of deformation  the Boussinesq solution is available \cite{Timoshenko}.   If the diameter of the circle is $l$ and the displacement is $\delta$, the elastic  energy reads \cite{Timoshenko},
$$
    \label{Busdef} \delta E_{B}= \frac{Y}{4(1-\nu^2)}l\delta^2. 
    $$
For the case of synchronic fluctuations, the Boussinesq deformations of the two bodies in contact will correspond, respectively to $+\delta$ and $-\delta$, leading to the factor $2$ for the total energy, which will be $2 \delta E_B$. 
The energy of this fluctuation should be of the order of the thermal energy, $k_BT$,  that is, $2\delta E_{B} \sim k_B T $, yielding the estimate, 
$$
l \sim \frac{4 k_B T(1-\nu^2)}{2 Y\delta^2 }.
$$
With $Y=120\cdot 10^9$N/m and $\nu =0.35$ for copper and the above values of $\tau $ and $\delta$, we obtain $l \sim 1.5 - 5$\,\AA, and respectively, the threshold velocity, 
$$
V_* \sim \frac{l}{\tau} \sim 0.15 - 1 \, \text{\AA/ps}, 
$$
with the median value of about $V_* \approx 0.5$\, \AA/ps\, =\, 50\, m/s.

One can also consider another form of the fluctuation, with the characteristic amplitude $\delta$ and characteristic length $l$. For instance the Hertzian deformation, for which an analytical result  for the energy is also available \cite{Timoshenko}. The respective estimates give, however, approximately the same result for the threshold velocity.

\subsection{Impact of non-synchronic fluctuations}
Using the approach sketched in the previous section, it is also possible to assess the impact of  non-synchronic fluctuations, when the deviation from the plane $z=0$ of the first body is $\delta u_z(x,y)$, and of the second body  is  $-\delta u_z(x,y)+ w_z(x,y)$. The obvious condition is $w_z(x,y) \geq 0$, that is, for  $w_z(x,y) =0$ the bodies remain in a tight contact, while for $w_z(x,y) > 0$ the contact is lost. The energy of such fluctuations is significantly  larger, since it contains linear with respect to $w_{z}(x,y)$ terms. Moreover, when the contact is lost, the free interface of the two bodies emerges, which requires an additional energy. For these reasons we expect that the main contribution  to the surface corrugation may be attributed to the synchronic surface fluctuations. Therefore we believe that our analysis made for the idealized case of synchronic surface fluctuations for identical bodies in a contact reflects the most prominent features of the phenomenon. The quantitative analysis of the impact of the non-synchronic fluctuations will be performed in future studies, where we will also develop the respective theory for a frictional contact of two bodies which are not identical.